\documentclass[onecolumn,secnumarabic,amsmath,amssymb,balancelastpage,nofootinbib]{article}

\usepackage[e]{esvect}
\usepackage{color}         
\usepackage{graphics}      
\usepackage{graphicx}      
\usepackage{epsf}          
\usepackage{bm}            
                                     
\usepackage{amssymb}
\usepackage{amsmath, esint}
\usepackage{mathrsfs}
\usepackage{framed}
\usepackage{bigints} 
\usepackage{enumitem}
\usepackage{libertine}
\usepackage[capbesideposition={left,center},facing=yes,capbesidewidth=8cm,capbesidesep=quad]{floatrow} 
\usepackage{setspace} 

\usepackage[none]{hyphenat} 

\usepackage[colorlinks=true]{hyperref}  

\definecolor{darkred}{rgb}{0.6,0,0}
\definecolor{darkgreen}{rgb}{0,0.5,0}
\definecolor{darkblue}{rgb}{0,0,0.6}
\hypersetup{ colorlinks,
linkcolor=darkred,
filecolor=darkgreen,
urlcolor=darkgreen,
citecolor=darkred }

\usepackage[style=apa, bibstyle=authortitle,labelyear=true,backend=biber,isbn=false,url=false,doi=true,eprint=false]{biblatex}
\AtEveryBibitem{\clearlist{language}}
\AtEveryBibitem{\clearlist{month}}
\newcommand{\citelink}[2]{\hyperlink{cite.\therefsection @#1}{#2}}

\renewbibmacro*{issue+date}{%
  \printtext[parens]{%
    \printfield{issue}%
    \setunit*{\addspace}%
    \usebibmacro{date}}%
  \newunit}

\title{\vspace{-4em} Relativistic Locality from Electromagnetism to Quantum Field Theory}
\author{%
\begin{tabular}{c} Eugene Y. S. Chua \\ Nanyang Technological University, Singapore\\ eugene.chuays@ntu.edu.sg \end{tabular} \and
\begin{tabular}{c} Charles T. Sebens \\ California Institute of Technology \\ csebens@caltech.edu \end{tabular}  }
\date{June 12, 2026 Preprint\\Forthcoming in \emph{Local Quantum Mechanics: \\ Everett, Many Worlds, and Reality} (ed.\ Alyssa Ney), \\ New York: Oxford University Press (2026)}

\begin{document}
\maketitle

\begin{abstract}
  \noindent
  Electromagnetism is the paradigm case of a theory that satisfies relativistic locality.  This can be proven by demonstrating that, once the theory’s laws are imposed, what is happening within a region fixes what will happen in the contracting light-cone with that region as its base.  The Klein-Gordon and Dirac equations meet the same standard.  We show that this standard can also be applied to quantum field theory (without collapse), examining two different ways of assigning reduced density matrix states to regions of space.  Our preferred method begins from field wave functionals and judges quantum field theory to be local.  Another method begins from particle wave functions (states in Fock space) and leads to either non-locality or an inability to assign states to regions, depending on the choice of creation operators.  We take this analysis of quantum field theory (without collapse) to show that the many-worlds interpretation of quantum physics is local at the fundamental level.  Debates about measurement and the branching of worlds do not challenge this fundamental locality.
\end{abstract}

\newpage
\begingroup
\tableofcontents
\endgroup

\newpage
\section{Introduction}

There is a well-known tension between special relativity and quantum physics.  Special relativity prohibits any instantaneous action at a distance and, more generally, any interaction across space-like separation.  This is the requirement of relativistic locality.  Quantum physics, by contrast, appears to involve just this sort of action at a distance.  The EPR argument and Bell’s theorem together show (from certain assumptions) that any theory that hopes to make the right probabilistic predictions (in cases like spin measurements of entangled particles located far apart from one another) must violate relativistic locality.  One can respond by embracing this non-locality and adopting a version of quantum physics that explicitly describes the non-local dynamics (as is done in Bohmian approaches).  Alternatively, one might hope to resolve the tension by abandoning one of the assumptions that goes into the EPR-plus-Bell argument for non-locality.  \textcite{maudlin_what_2014} highlights the assumption that measurements have unique results, noting that the argument does not directly apply to the many-worlds interpretation of quantum mechanics (where every possible result occurs on some branch).  He then writes:
\begin{quote}
``That does not prove that Many Worlds is local: it just shows that Bell’s result does not prove that it isn’t local. In order to even address the question of the locality of Many Worlds a tremendous amount of interpretive work has to be done. This is not the place to attempt such a task.’’ (\cite[pg.\ 23]{maudlin_what_2014})\footnote{See \textcite[sec.\ 10.4]{norsen_foundations_2017} for more on this challenge.}
\end{quote}
Here we attempt that task, arguing that the many-worlds interpretation is indeed local by analyzing quantum field theory (QFT) in a ``many-worlds’’ or ``Everettian’’ form, where there is no true collapse of the quantum state and no supplementation of that state with additional ontology (like point particles with definite locations or fields with definite configurations).  Determining whether the collapse-free Schr\"{o}dinger dynamics in QFT is local is important for accurately assessing the merits of the many-worlds interpretation, and also for better understanding QFT independently of that interpretation.

In section \ref{EMsection}, we begin with classical electromagnetism.  Electromagnetism is a local theory because once you have specified what is happening in a spherical region, the laws of the theory only allow a single possible future in the contracting light-cone that has that region as its base.  Put another way: the future of that spherical region is fixed within a zone that shrinks at the speed of light.  This is a standard for relativistic locality that can be applied across many domains of physics.

In section \ref{KGDsection}, we see that the Klein-Gordon and Dirac equations meet the same standard.  In section \ref{KGsection}, we prove this for the Klein-Gordon equation.  In section \ref{SPsection}, we address worries about superluminal propagation that have been raised for the Klein-Gordon equation.  In section \ref{DEsection}, we offer a proof of locality for the Dirac equation that closely mirrors the proofs of locality for electromagnetism and the Klein-Gordon equation (something that we have not seen done before).

We turn to QFT in section \ref{QFTsection}.  To apply the standard for relativistic locality from section \ref{EMsection}, we must first find a way to specify the state of a region of space at a time.  \textcite{wallace_quantum_2010} propose a general way of assigning states to spacetime regions and sketch how locality can be proved for such states.  We aim to add fine lines to their broad brushstrokes, and consider two strategies for making the proposal more ontologically precise.  First, one could represent the universal state by a field wave functional and construct an appropriate field reduced density matrix for a region by tracing over the field degrees of freedom outside the region.  Second, one could represent the universal state by a particle wave function for a variable number of particles (a state in Fock space) and attempt to construct an appropriate particle reduced density matrix for a region by tracing over the particle degrees of freedom outside the region.  We explore both options in detail and show, at least in outline, how relativistic locality can be proven within a field approach but \textit{not} a particle approach (where the standard creation operators prevent one from assigning states to regions, and the alternative Newton-Wigner creation operators lead to non-locality).  The locality of the many-worlds interpretation thus appears to turn on the choice between particle and field approaches.  We prefer the field wave functional approach for a variety of reasons and see its locality as another point in its favor.

In the conclusion, we briefly discuss the non-fundamental carving of the quantum state into distinct worlds.  The debate as to how such worlds branch (whether locally or globally) does not threaten our argument that, at the fundamental level, the many-worlds interpretation is local.

\section{Electromagnetism and the Standard for Relativistic Locality}\label{EMsection}

Electromagnetism sets the gold standard for relativistic locality.  It is widely understood to be a local theory where causal influences cannot propagate faster than the speed of light.  Let us now examine how that can be proven (employing Gaussian CGS units).\footnote{\textcite[sec.\ 10.2]{wald_general_1984}; \textcite[sec.\ 5.4]{wald_advanced_2022} presents a slightly different way of proving the same result.}

Let us begin by studying the behavior of the vector and scalar potentials $\vec{A}$ and $\phi$, related to the electric and magnetic fields via
\begin{align}
\vec{E}&=-\vec{\nabla}\phi-\frac{1}{c}\frac{\partial \vec{A}}{\partial t}
\nonumber
\\
\vec{B}&=\vec{\nabla} \times \vec{A}
\ ,
\label{fieldsfrompotentials}
\end{align}
adopting the Lorenz gauge as a partial gauge fixing condition,
\begin{equation}
\vec{\nabla}\cdot\vec{A}+\frac{1}{c}\frac{\partial \phi}{\partial t}=0
\ .
\label{lorenz}
\end{equation}
Using potentials that satisfy \eqref{fieldsfrompotentials} ensures that two of Maxwell's equations are automatically satisfied.  In the Lorenz gauge, the other two of Maxwell's equations become wave equations for the potentials, with the charge and current densities acting as source terms,
\begin{align}
\bigg(\nabla^2-\frac{1}{c^2}\frac{\partial^2}{\partial t^2}\bigg)\phi&=-4\pi\rho
\label{phiwave}
\\ 
\bigg(\nabla^2-\frac{1}{c^2}\frac{\partial^2}{\partial t^2}\bigg)\vec{A}&=-\frac{4\pi}{c}\vec{J}
\ .
\label{awave}
\end{align}

To prove locality for a complete theory where the electromagnetic field interacts with charged matter, we would also need to introduce equations describing the way that the charge and current densities change in response to the electromagnetic field---and use those equations to show that $\phi$, $\vec{A}$, $\rho$, and $\vec{J}$ evolve together in a local way.  For our purposes in this section, we would like to leave the type of matter and its response to electromagnetic fields unspecified.  So, we will follow standard practice and treat $\rho$ and $\vec{J}$ as fixed source functions given at every point in space and time (obeying the equation for local conservation of charge, $\frac{\partial \rho}{\partial t} = -\vec{\nabla}\cdot \vec{J}\,$).  The goal then is to show that, for specified sources, the above wave equations yield a local evolution for the potentials.

\begin{figure}[h!]
\center{\includegraphics[width=8 cm]{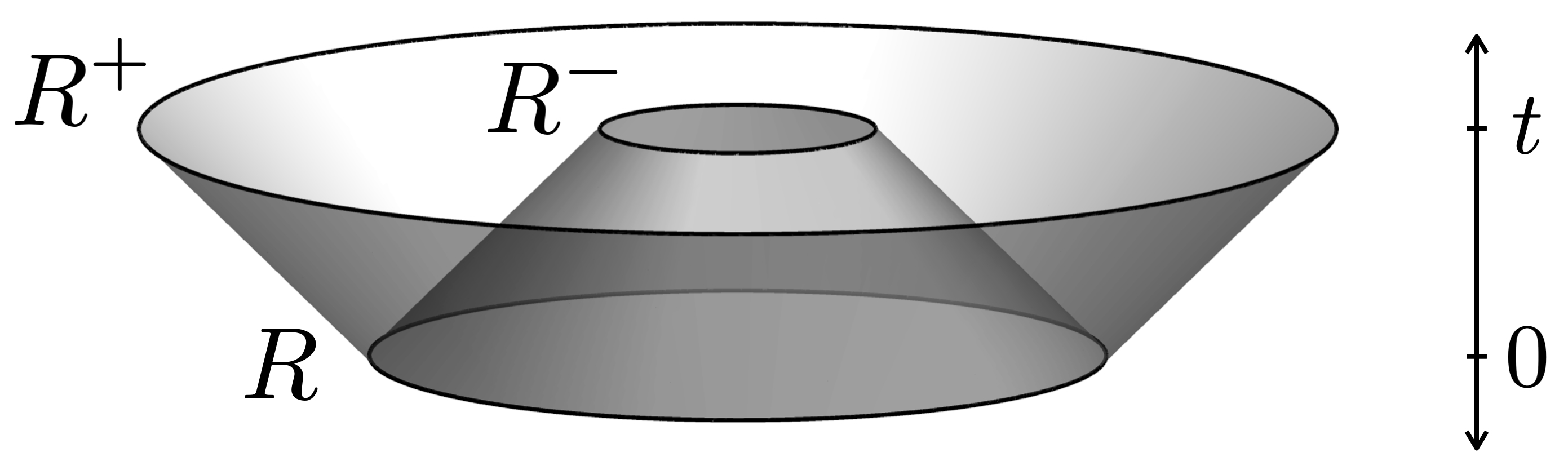}}
\caption{This figure shows expanding and contracting light-cones from $R$ at $t=0$, picking out the regions $R^+$ and $R^-$ at a later time $t$.  In a local theory, what is happening within $R$ at $t=0$ should at that later time fully determine what happens within $R^-$ and only influence what happens within $R^+$.}
  \label{lightcones}
\end{figure}

For electromagnetism (formulated in terms of potentials in the Lorenz gauge) to be relativistically local, it must be the case that specifying what is happening inside a spherical region $R$, at some time that we can call $t=0$, fixes what happens everywhere inside the contracting light-cone with $R$ as its base.  Put more precisely, any two solutions to the wave equations that agree on the values of $\phi$, $\frac{\partial \phi}{\partial t}$, $\vec{A}$, and $\frac{\partial \vec{A}}{\partial t}$ within $R$ at $t=0$ must agree on the values of $\phi$ and $\vec{A}$ within the entire contracting light-cone with $R$ as its base (given the fixed source terms specified throughout the light-cone).  This standard of relativistic locality asks what $R$ determines in the future.  We can alternatively phrase it as a standard about what $R$ can influence: any two solutions to the wave equations that disagree on the values of $\phi$, $\frac{\partial \phi}{\partial t}$, $\vec{A}$, and $\frac{\partial \vec{A}}{\partial t}$ only within $R$ at $t=0$ can later disagree on the values of $\phi$ and $\vec{A}$ only within the expanding light-cone that has $R$ as its base (given the fixed source terms specified throughout the light-cone).  See figure \ref{lightcones}.

The next steps largely follow standard proofs from partial differential equations textbooks that three-dimensional wave equations like \eqref{phiwave} and \eqref{awave} satisfy a ``principle of causality'' (another name for relativistic locality).\footnote{Our proof will most closely parallel \textcite[sec.\ 9.1]{strauss_partial_2008}, straightforwardly generalized to include source terms.  Similar proofs appear in \textcite[ch.\ 8]{zachmanoglou_introduction_1976}, \textcite[ch.\ 5]{folland_introduction_1995}, and \textcite[sec.\ 2.4.3]{evans_partial_1998}.}  These steps can be skipped or skimmed on a first reading of this paper, but they provide a useful illustration of what it takes to prove locality, and it is by modifying this proof that we will arrive at proofs of locality for the Klein-Gordon and Dirac equations.

\begin{figure}[h!]
\center{\includegraphics[width=13.5 cm]{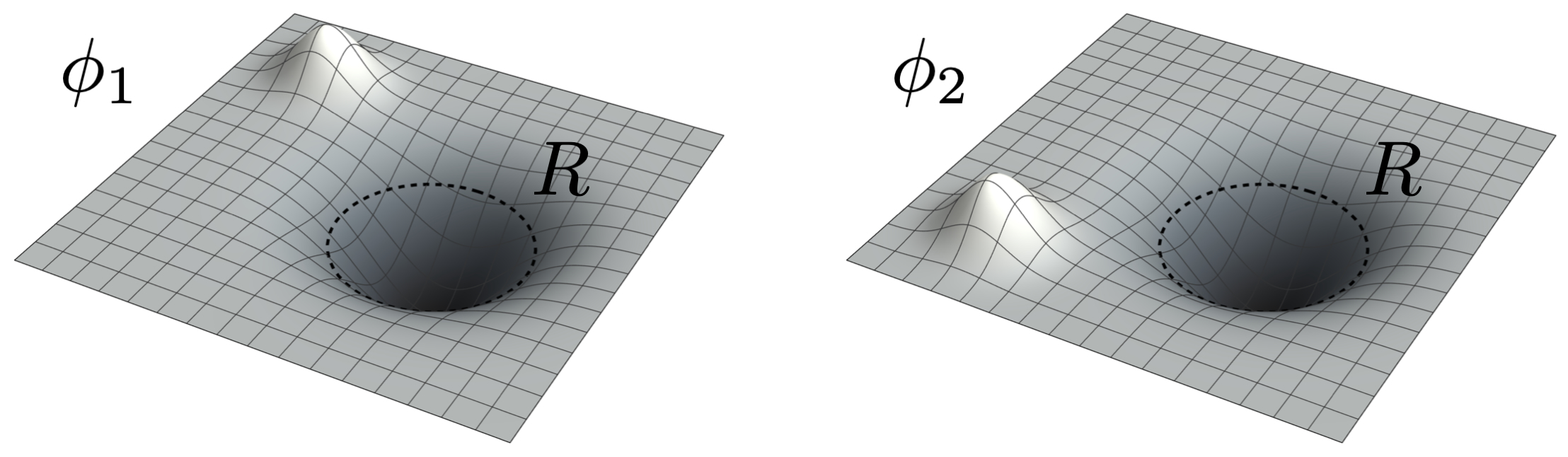}}
\caption{Scalar potentials $\phi_1$ and $\phi_2$ that initially only differ outside of $R$ will agree within the contracting light-cone of $R$.}
  \label{potentials}
\end{figure}

To show that what is happening in the sphere $R$ at $t=0$ fixes what happens within the entire contracting light-cone, let us select an arbitrary time $t$ (after $t=0$ but before the contracting light-cone vanishes), at which point the contracting light-cone from $R$ has shrunk to the smaller sphere $R^-$, and show that the state on $R^-$ is determined uniquely.  Let us focus first on the scalar potential $\phi$ and its wave equation \eqref{phiwave}.  Consider two separate solutions to \eqref{phiwave}, $\phi_1(\vec{x},t)$ and $\phi_2(\vec{x},t)$, that agree on their values and their first time derivatives within $R$ at $t=0$ (figure \ref{potentials}).  What we would like to show is that these solutions must agree within $R^-$ at $t$.  To do so, let us begin by defining their difference as
\begin{equation}
\phi_d=\phi_1-\phi_2
\ .
\end{equation}
If $\phi_1$ and $\phi_2$ obey the wave equation \eqref{phiwave} for a specified source $\rho$, then $\phi_d$ must obey the source-free (homogeneous) wave equation:
\begin{align}
\bigg(\nabla^2-\frac{1}{c^2}\frac{\partial^2}{\partial t^2}\bigg)\phi_d&=0
\ .
\label{phidwave}
\end{align}
Multiplying through by $-c^2\frac{\partial \phi_d}{\partial t}$ and rearranging yields
\begin{align}
0&=-c^2\frac{\partial \phi_d}{\partial t}\bigg(\nabla^2-\frac{1}{c^2}\frac{\partial^2}{\partial t^2}\bigg)\phi_d
\nonumber
\\
&=\frac{\partial \phi_d}{\partial t}\frac{\partial^2\phi_d}{\partial t^2}+\frac{c^2}{2}\frac{\partial}{\partial t}|\vec{\nabla}\phi_d|^2-c^2\vec{\nabla}\cdot\bigg(\frac{\partial \phi_d}{\partial t}\vec{\nabla}\phi_d\bigg)
\nonumber
\\
&=\frac{\partial}{\partial t}\left[\frac{1}{2}\bigg(\frac{\partial \phi_d}{\partial t}\bigg)^2+\frac{c^2}{2}|\vec{\nabla}\phi_d|^2\right]-c^2\vec{\nabla}\cdot\bigg(\frac{\partial \phi_d}{\partial t}\vec{\nabla}\phi_d\bigg)
\ .
\label{phidwave2}
\end{align}

\begin{figure}[htb]
\center{\includegraphics[width=4 cm]{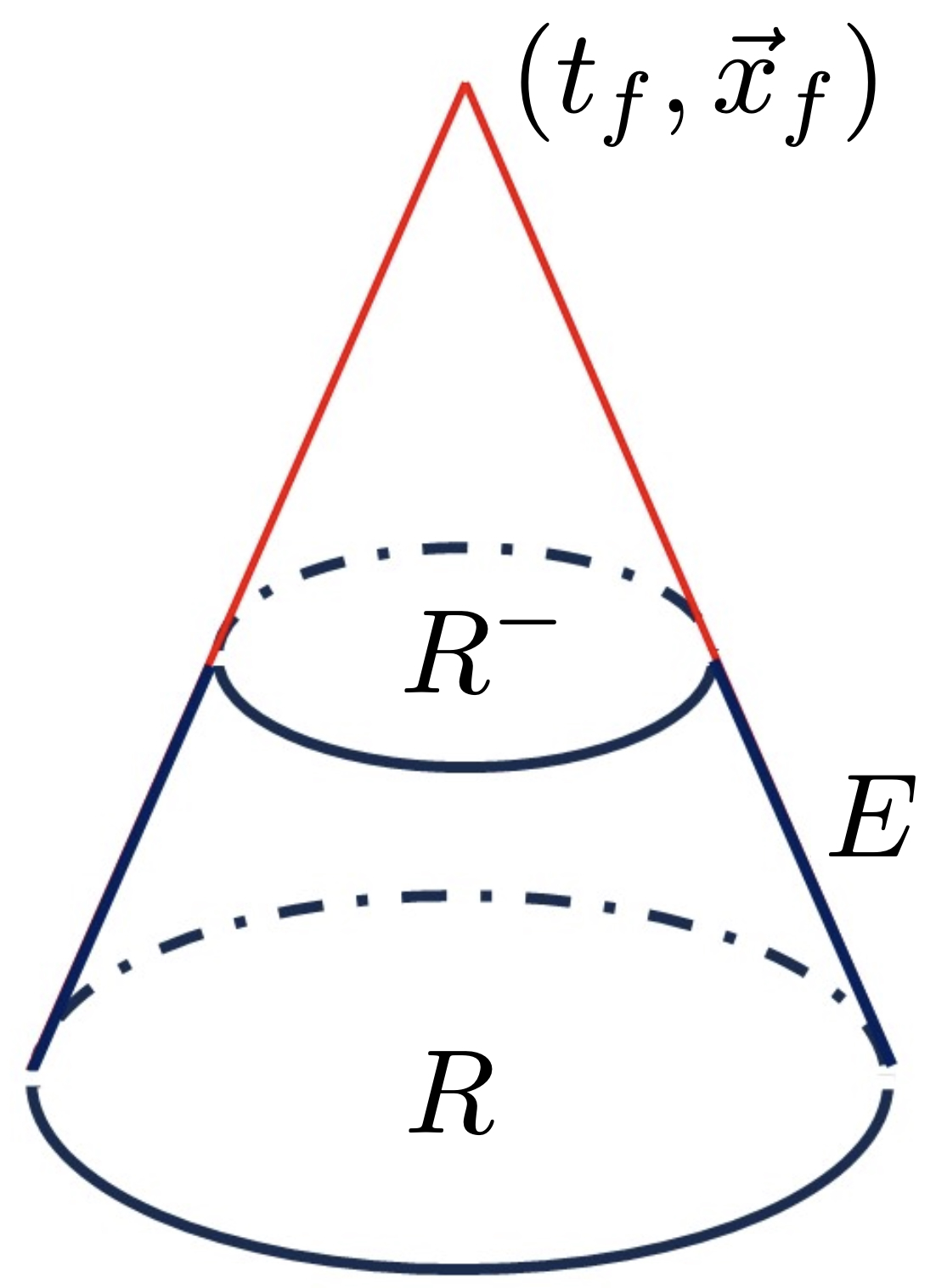}}
\caption{The frustum $F$ is the portion of the light-cone bounded by the base, $R$, the top, $R^-$, and the edge, $E$.}
  \label{frustumfig}
\end{figure}

To show that $\phi_d$ vanishes on $R^-$, it will turn out to be useful to integrate the above expression over the solid portion $F$ of the light-cone that has $R$ at its base, $R^-$ as its top, and $E$ as its edge (figure \ref{frustumfig}).  (This shape is called a ``frustum,'' hence the capital $F$.)  The integral over the frustum $F$ can be transformed (using the four-dimensional divergence theorem) from a four-dimensional integral over the entirety of $F$ to an integral over the three-dimensional ``surfaces'' of $F$ ($R$, $R^-$, and $E$),
\begin{align}
0&=\int_F d^3\vec{x} dt \bigg(\frac{\partial}{\partial t}\left[\frac{1}{2}\bigg(\frac{\partial \phi_d}{\partial t}\bigg)^2+\frac{c^2}{2}|\vec{\nabla}\phi_d|^2\right]-c^2\vec{\nabla}\cdot\bigg(\frac{\partial \phi_d}{\partial t}\vec{\nabla}\phi_d\bigg)\bigg)
\nonumber
\\
&=\int_{R\cup R^- \cup E} d^3\vec{x} \bigg(n_t \left[\frac{1}{2}\bigg(\frac{\partial \phi_d}{\partial t}\bigg)^2+\frac{c^2}{2}|\vec{\nabla}\phi_d|^2\right]-\vec{n}\cdot\bigg(c^2\frac{\partial \phi_d}{\partial t}\vec{\nabla}\phi_d\bigg)\bigg)
\ .
\label{phidwave3}
\end{align}
Here, $(n_t,\vec{n})$ is the unit normal four-vector pointing outward from $F$ on the three-dimensional boundary of $F$.  For the bottom of the frustum, $R$, the unit normal four-vector points temporally downward $(-1,0)$ and for the top, $R^-$, this points temporally upwards $(1,0)$.  The integral can thus be broken up into the following three contributions:
\begin{align}
0&=-\int_{R} d^3\vec{x} \bigg(\frac{1}{2}\bigg(\frac{\partial \phi_d}{\partial t}\bigg)^2+\frac{c^2}{2}|\vec{\nabla}\phi_d|^2\bigg)
\nonumber
\\
&\quad\quad +\int_{R^-} d^3\vec{x} \bigg(\frac{1}{2}\bigg(\frac{\partial \phi_d}{\partial t}\bigg)^2+\frac{c^2}{2}|\vec{\nabla}\phi_d|^2\bigg)
\nonumber
\\
&\quad\quad\quad\quad +\int_{E} d^3\vec{x} \bigg(n_t \left[\frac{1}{2}\bigg(\frac{\partial \phi_d}{\partial t}\bigg)^2+\frac{c^2}{2}|\vec{\nabla}\phi_d|^2\right]-\vec{n}\cdot\bigg(c^2\frac{\partial \phi_d}{\partial t}\vec{\nabla}\phi_d\bigg)\bigg)
\ .
\label{phidwaveint}
\end{align}
The first integral vanishes because we assumed earlier that the two solutions $\phi_1$ and $\phi_2$ agreed on what was happening within $R$ at $t=0$: $\phi_1=\phi_2$ and $\frac{\partial \phi_1}{\partial t}=\frac{\partial \phi_2}{\partial t}$, thus $\phi_d=\frac{\partial\phi_d}{\partial t}=0$.  We are left with the sum of the integrals over $R^-$ and $E$ being zero,
\begin{align}
&\int_{R^-} d^3\vec{x} \bigg(\frac{1}{2}\bigg(\frac{\partial \phi_d}{\partial t}\bigg)^2+\frac{c^2}{2}|\vec{\nabla}\phi_d|^2\bigg)
\nonumber
\\
&\quad\quad +\int_{E} d^3\vec{x} \bigg(n_t \left[\frac{1}{2}\bigg(\frac{\partial \phi_d}{\partial t}\bigg)^2+\frac{c^2}{2}|\vec{\nabla}\phi_d|^2\right]-\vec{n}\cdot\bigg(c^2\frac{\partial \phi_d}{\partial t}\vec{\nabla}\phi_d\bigg)\bigg)=0
\ .
\label{phidwaveint2}
\end{align}

Letting $(t_f,\vec{x}_f)$ denote the tip of the light-cone (floating above the frustum in figure \ref{frustumfig}), the unit outward normal four-vector on $E$ can be written as\footnote{See \textcite[pg.\ 229, 232]{strauss_partial_2008}.}
\begin{equation}
(n_t,\vec{n}) = \frac{c}{\sqrt{c^2+1}} \bigg(1, \frac{\vec{x} - \vec{x}_f}{c|\vec{x} - \vec{x}_f|}\bigg)
\ .
\label{4Dnormal}
\end{equation}
Inserting this four-vector, the integral over $E$ in \eqref{phidwaveint2} becomes
\begin{align}
&\int_{E} d^3\vec{x} \bigg(n_t \left[\frac{1}{2}\bigg(\frac{\partial \phi_d}{\partial t}\bigg)^2+\frac{c^2}{2}|\vec{\nabla}\phi_d|^2\right]-\vec{n}\cdot\bigg(c^2\frac{\partial \phi_d}{\partial t}\vec{\nabla}\phi_d\bigg)\bigg)
\nonumber
\\
&\quad\quad=\frac{c}{\sqrt{c^2+1}}\int_{E} d^3\vec{x} \bigg(\frac{1}{2}\bigg(\frac{\partial \phi_d}{\partial t}\bigg)^2+\frac{c^2}{2}|\vec{\nabla}\phi_d|^2-c^2\frac{\partial \phi_d}{\partial t}\frac{\vec{x} - \vec{x}_f}{c|\vec{x} - \vec{x}_f|}\cdot\vec{\nabla}\phi_d\bigg)
\ .
\label{Eint}
\end{align}
This can equivalently be written as
\begin{align}
&\frac{c}{\sqrt{c^2+1}}\int_{E} d^3\vec{x} \left[\frac{1}{2}\bigg(\frac{\partial \phi_d}{\partial t}-c\frac{\vec{x} - \vec{x}_f}{|\vec{x} - \vec{x}_f|}\cdot\vec{\nabla}\phi_d\bigg)^2+\frac{c^2}{2}\bigg(\vec{\nabla} \phi_d -\bigg( \frac{\vec{x} - \vec{x}_f}{|\vec{x} - \vec{x}_f|}\cdot\vec{\nabla}\phi_d\bigg) \frac{\vec{x} - \vec{x}_f}{|\vec{x} - \vec{x}_f|}\bigg)^2\right]
\ .
\label{Eint2}
\end{align}
The equivalence of \eqref{Eint} and \eqref{Eint2} can be checked by expanding the two squared expressions in \eqref{Eint2}.

Because \eqref{Eint2} is an integral of two squares that each must make a non-negative contribution, it must be greater than or equal to zero.  Returning to \eqref{phidwaveint2}, this means that the integral over $R^-$ must be less than or equal to zero,
\begin{align}
&\int_{R^-} d^3\vec{x} \bigg(\frac{1}{2}\bigg(\frac{\partial \phi_d}{\partial t}\bigg)^2+\frac{c^2}{2}|\vec{\nabla}\phi_d|^2\bigg) \leq 0
\ .
\label{finalinequality}
\end{align}
In this integral, $\big(\frac{\partial \phi_d}{\partial t}\big)^2$ and $\frac{c^2}{2}|\vec{\nabla}\phi_d|^2$ are squares that are non-negative everywhere.  So, the $\leq$ relation can only be satisfied if $\frac{\partial \phi_d}{\partial t}$ and $\vec{\nabla}\phi_d$ both vanish on the entirety of $R^-$.  Because $t$ was chosen arbitrarily in picking out this particular frustum, $\frac{\partial \phi_d}{\partial t}$ and $\vec{\nabla}\phi_d$ must be zero throughout the contracting light-cone.  As $\phi_d=0$ at the base of the light-cone $R$, it thus must be zero throughout the light-cone.  That is, the two solutions $\phi_1$ and $\phi_2$ that agreed on what was happening in $R$ must agree throughout the contracting light-cone.  This completes the proof of locality for the scalar potential $\phi$.

Because the wave equations for each component of the vector potential $\vec{A}$ take the exact same form as the wave equation for $\phi$, the very same techniques as those employed above can be used to prove that the evolution of $\vec{A}$ in the Lorenz gauge is local.  If two solutions to \eqref{awave} agree on $\vec{A}$ and $\frac{\partial \vec{A}}{\partial t}$ within $R$ at $t=0$, they will agree throughout the contracting light-cone.  Looking at $\phi$ and $\vec{A}$ together: a specification of $\phi$, $\frac{\partial \phi}{\partial t}$, $\vec{A}$, and $\frac{\partial \vec{A}}{\partial t}$ on $R$---satisfying the Lorenz gauge condition \eqref{lorenz}---determines uniquely the evolution of $\phi$ and $\vec{A}$ within the contracting light-cone.

If you take the fundamental ontology of electromagnetism to be the scalar and vector potentials in the Lorenz gauge, then our proof of locality could end here.  The dynamical laws, \eqref{phiwave} and \eqref{awave}, are second-order and thus you need to know both the values of the potentials and their derivatives (along with any source terms) to generate time evolution.  Specifying these quantities within a sphere fully determines what will happen in the contracting light-cone with that sphere as its base.  If, on the other hand, you take the fundamental ontology to be the gauge-invariant electric and magnetic fields, then there are a few more steps to prove locality.\footnote{See \textcite[pg.\ 96--97]{wald_advanced_2022}.}  The dynamical equations for these fields (the two of Maxwell's equations that feature derivatives of the fields) are first-order and thus you need to know only the values of the electric and magnetic fields at a moment (along with any source terms) to generate time evolution.  To prove locality, we must show that a specification of $\vec{E}$ and $\vec{B}$ within a sphere $R$ fixes what happens in the contracting light-cone (assuming that the source terms are given).

Let us begin by drawing a relation between two different ways of presenting the initial conditions: It is a fact that specifying the values of $\vec{E}$ and $\vec{B}$ within a sphere $R$ at $t=0$ (satisfying Maxwell's equations $\vec{\nabla}\cdot\vec{B}=0$ and $\vec{\nabla} \cdot \vec{E}=4\pi\rho$) will fix $\phi$, $\frac{\partial \phi}{\partial t}$, $\vec{A}$, and $\frac{\partial \vec{A}}{\partial t}$ in the Lorenz gauge on $R$, up to the gauge transformation that remains because the Lorenz gauge is only a partial gauge fixing condition, leaving the freedom to transform
\begin{align}
\phi &\rightarrow \phi-\frac{\partial \Lambda}{\partial t}
\nonumber
\\
\vec{A} &\rightarrow \vec{A} + \vec{\nabla}\Lambda
\label{LGtransf}
\end{align}
for any scalar field $\Lambda$ such that $\nabla^2 \Lambda-\frac{1}{c^2}\frac{\partial \Lambda}{\partial t^2}=0$.\footnote{To see that this fact holds, note that (a) $\vec{B}=\vec{\nabla} \times \vec{A}$ specifies $\vec{A}$ up to the gauge transformation $\vec{A} \rightarrow \vec{A} + \vec{\nabla}\Lambda$; (b) because $\vec{B}=\vec{\nabla} \times \vec{A}$ holds at all times, $\frac{\partial \vec{A}}{\partial t}$ must transform as $\frac{\partial \vec{A}}{\partial t} \rightarrow \frac{\partial \vec{A}}{\partial t} + \frac{\partial}{\partial t}\vec{\nabla}\Lambda$; (c) given that $\frac{\partial \vec{A}}{\partial t}$ transforms this way, $\vec{E}=-\vec{\nabla}\phi-\frac{1}{c}\frac{\partial \vec{A}}{\partial t}$ allows us to determine $\phi$ and $\frac{\partial \vec{A}}{\partial t}$ up to the gauge transformations $\phi \rightarrow \phi-\frac{\partial \Lambda}{\partial t}$ and $\frac{\partial \vec{A}}{\partial t} \rightarrow \frac{\partial \vec{A}}{\partial t} + \frac{\partial}{\partial t}\vec{\nabla}\Lambda$; (d) for Gauss's law $\vec{\nabla} \cdot \vec{E}=4\pi\rho$ to be satisfied in the Lorenz gauge, it must be the case that $\nabla^2 \Lambda-\frac{1}{c^2}\frac{\partial \Lambda}{\partial t^2}=0$ as at the end of \eqref{LGtransf}; (e) using this relation, the Lorenz gauge condition \eqref{lorenz} allows one to go from the specification of $\vec{A}$ (up to a gauge transformation) from part (a) to a determination of $\frac{\partial \phi}{\partial t}$ up to the gauge transformation $\frac{\partial \phi}{\partial t} \rightarrow \frac{\partial \phi}{\partial t}-\frac{\partial^2 \Lambda}{\partial t^2}$.}  From the given $\vec{E}$ and $\vec{B}$ fields on $R$, let us arbitrarily pick one set of $\phi$, $\frac{\partial \phi}{\partial t}$, $\vec{A}$, and $\frac{\partial \vec{A}}{\partial t}$ in the Lorenz gauge to encode the fields among those sets allowed by the remaining gauge freedom \eqref{LGtransf}. Then, our earlier proof shows that there is only one possible evolution of these potentials within the contracting light-cone.  By \eqref{fieldsfrompotentials}, this means that there is only one possible evolution for $\vec{E}$ and $\vec{B}$ within the contracting light-cone. Other choices for the potentials that differ by the gauge transformations in \eqref{LGtransf} must yield the same dynamics for $\vec{E}$ and $\vec{B}$ because they will only ever differ by a gauge transformation.  We have thus proven locality for the dynamics of the $\vec{E}$ and $\vec{B}$ fields.

 Before moving on, let us carefully state the general standard for locality (in deterministic theories) that we have applied in this section.\footnote{Relativistic locality can be contrasted with the related concept of spatiotemporal locality or continuous action (\cite[pg.\ 13--17]{lange_introduction_2002}; \cite{wharton_colloquium_2020}).  A theory is spatiotemporally local if interactions occur without any gaps in space or time. This standard allows influences to travel arbitrary quickly, provided they do not jump any gaps when doing so.  By contrast, relativistic locality is a stricter standard that enforces a light-speed limit on interactions.  (Adlam, \citelink{adlam_should_nodate}{this volume}, introduces a standard for continuous action that includes a light-speed limit and should be satisfied by theories that meet our standard for relativistic locality.  Such theories should also meet her condition prohibiting action at a distance.)}
 \begin{quote}
     \textbf{The Standard for Relativistic Locality:}\\
     A deterministic theory with \textbf{first-order dynamics} is \textbf{relativistically local} if and only if, given the theory's laws, the instantaneous physical state within a sphere $R$ at some time uniquely determines, at any future time, the instantaneous physical state within the smaller sphere $R^-$ that is a slice of the contracting light-cone with $R$ as its base (figure \ref{lightcones}). \\
     A deterministic theory with \textbf{second-order dynamics} is \textbf{relativistically local} if and only if, given the theory's laws, the instantaneous physical state \emph{and rates of change} within a sphere $R$ at some time uniquely determines, at any future time, the instantaneous physical state within the smaller sphere $R^-$ that is a slice of the contracting light-cone with $R$ as its base (figure \ref{lightcones}).
 \end{quote}
Here the slogan that ``what is happening in $R$'' should fix what happens in its contracting light-cone gets filled in differently depending on whether we are dealing with first-order or second-order dynamics.  For first-order dynamics (like Maxwell's equations for the evolution of the electric and magnetic fields), all that needs to be specified is the instantaneous physical state at $t=0$ (such as $\vec{E}$ and $\vec{B}$).  For second-order dynamics (like the wave equations for the potentials in the Lorenz gauge), we must specify the instantaneous physical state and rates of change at $t=0$ (such as $\phi$, $\frac{\partial \phi}{\partial t}$, $\vec{A}$, and $\frac{\partial \vec{A}}{\partial t}$).

The above standard is applied within a particular reference frame, but it is not frame-dependent.  In a relativistic theory the same laws hold in all inertial frames, and thus if the condition is met in one inertial frame it is met in all inertial frames.

The above standard is appropriate for deterministic theories and might be called a standard of ``local determinism.''  For theories where there is randomness in the dynamics, one would not expect specifying what is initially happening within $R$ to uniquely determine the state of $R^-$.  Such a theory would be relativistically local if specifying what is initially happening within $R$ gives a probability distribution over different states of $R^-$ that is independent of what is initially happening outside of $R$.

The standard for relativistic locality that we have put forward focuses on the contracting light-cone of a region.  We could alternatively (equivalently) look at the expanding light-cone of a region and ask whether two initial specifications of what is happening that only disagree within $R$ later only disagree within $R^+$ (figure \ref{lightcones}). Vaidman (\citelink{vaidman_many-worlds_nodate}{this volume}) formulates his similar standard (which he describes as a prohibition on action at a distance) this way.

\section{A Bridge Between Electromagnetism and Quantum Field Theory}\label{KGDsection}

As a bridge between the proof of locality that we have just seen for electromagnetism and the question of locality in QFT, this section discusses the Klein-Gordon and Dirac equations.   These equations are often presented as part of ``relativistic quantum mechanics,'' where they are seen as giving the dynamics for quantum wave functions (of particles with spin-$0$ or spin-$1/2$, respectively).  QFT can then potentially be arrived at by extending such single-particle relativistic equations to scenarios involving particle creation and annihilation.  Alternatively, the Klein-Gordon and Dirac equations are sometimes interpreted as field equations giving the dynamics of classical relativistic fields.  The path to QFT is then to quantize such classical fields.  The merits of these two routes to QFT are discussed in \textcite[ch. 11]{durr_understanding_2020}; \textcite{sebens_fundamentality_2022}; \textcite[sec.\ 6.5]{tumulka_foundations_2022}.  For our purposes here, we can analyze locality for the Klein-Gordon and Dirac equations while keeping in mind that they might be interpreted either as equations of relativistic quantum mechanics or relativistic classical field theory.  Studying these equations serves as a useful warm-up to QFT.

\subsection{The Klein-Gordon Equation}\label{KGsection}

It is straightforward to prove relativistic locality for the second-order Klein-Gordon equation in a way that parallels our earlier proof of relativistic locality for the wave equation governing the evolution of the electromagnetic scalar potential in the Lorenz gauge.\footnote{\textcite[pg.\ 234]{strauss_partial_2008} and \textcite[pg.\ 95]{wald_general_1984} both mention that their proofs of ``causality'' (locality) can be applied to the Klein-Gordon equation as well.}  Let us begin with the free Klein-Gordon equation,
\begin{equation}
\bigg(\frac{1}{c^2}\frac{\partial^2}{\partial t^2} - \nabla^2 + \frac{m^2 c^2}{\hbar^2}\bigg)\phi=0
\ ,
\label{KQeq}
\end{equation}
which differs from the wave equation for the scalar potential \eqref{phiwave} only by the addition of a mass term (and puts the symbol $\phi$ to a new use).  To prove locality, we must show that a specification of $\phi$ and $\frac{\partial \phi}{\partial t}$ within some sphere $R$ fixes what will happen within the contracting light-cone that has $R$ as its base (figure \ref{frustumfig}).  As in the earlier proof, we can consider two solutions $\phi_1$ and $\phi_2$ that agree on their values and their first time derivatives within $R$ at $t=0$. Their difference $\phi_d=\phi_1-\phi_2$ will also be a solution to the Klein-Gordon equation:
\begin{equation}
\bigg(\frac{1}{c^2}\frac{\partial^2}{\partial t^2} - \nabla^2 + \frac{m^2 c^2}{\hbar^2}\bigg)\phi_d=0
\ ,
\end{equation}
As before, we can multiply through by $c^2\frac{\partial \phi_d}{\partial t}$, rearrange, and integrate over the frustum $F$ to get
\begin{align}
0&=\int_F d^3\vec{x} dt \bigg(\frac{\partial}{\partial t}\left[\frac{1}{2}\bigg(\frac{\partial \phi_d}{\partial t}\bigg)^2+\frac{c^2}{2}|\vec{\nabla}\phi_d|^2+\frac{m c^4}{2 \hbar^2}\phi_d^2\right]-c^2\vec{\nabla}\cdot\bigg(\frac{\partial \phi_d}{\partial t}\vec{\nabla}\phi_d\bigg)\bigg)
\ ,
\end{align}
in parallel to \eqref{phidwave3}.  Using the four-dimensional divergence theorem, this can again be broken up into integrals over $R$, $R^-$ and $E$.  The integral over $R$ will again vanish because $\phi_d$ and $\frac{\partial \phi_d}{\partial t}$ are (by supposition) 0 on $R$ at $t=0$.  The integral over $E$ is as before \eqref{Eint2}, with an additional term that must make a positive contribution,
\begin{align}
&\frac{c}{\sqrt{c^2+1}}\int_{E} d^3\vec{x} \left[\frac{1}{2}\bigg(\frac{\partial \phi_d}{\partial t}-c\frac{\vec{x} - \vec{x}_f}{|\vec{x} - \vec{x}_f|}\cdot\vec{\nabla}\phi_d\bigg)^2+\frac{c^2}{2}\bigg(\vec{\nabla} \phi_d -\bigg( \frac{\vec{x} - \vec{x}_f}{|\vec{x} - \vec{x}_f|}\cdot\vec{\nabla}\phi_d\bigg) \frac{\vec{x} - \vec{x}_f}{|\vec{x} - \vec{x}_f|}\bigg)^2+\frac{m c^4}{2 \hbar^2}\phi_d^2\right]
\ .
\end{align}
So, again the integral over $E$ is $\geq 0$.  The integral over $R^-$ must then satisfy the inequality,
\begin{align}
&\int_{R^-} d^3\vec{x} \bigg(\frac{1}{2}\bigg(\frac{\partial \phi_d}{\partial t}\bigg)^2+\frac{c^2}{2}|\vec{\nabla}\phi_d|^2\bigg) \leq 0
\ ,
\end{align}
as in \eqref{finalinequality}.  This can only be achieved if $\frac{\partial \phi_d}{\partial t}$, $\vec{\nabla}\phi_d$, and $\phi_d$ all vanish on the entirety of $R^-$.  Because this must hold for any choice of $t$ and corresponding slice of the contracting light-cone $R^-$, $\phi_1$ and $\phi_2$ must agree throughout the contracting light-cone.  This establishes locality for the free Klein-Gordon equation.

\subsection{Interlude: Superluminal Propagation}\label{SPsection}

In the literature on probability densities and position operators for quantum wave functions obeying the Klein-Gordon equation, you sometimes see claims that certain states will propagate superluminally---in apparent contradiction with the proof of locality that we have just presented.  Let us now address that apparent contradiction to assuage concerns about superluminal propagation.

\textcite{pagonis_strange_1999} provide a way of dissolving the tension:
\begin{quote}
``\dots\ the restriction to subluminal propagation refers to wavefunctions for which \emph{both} $\phi$ and its time derivative $\partial \phi / \partial x^0$ have momentarily compact support. \dots\ there are no such wave functions in the positive (or indeed, the negative) energy subspace.''
(\cite[pg.\ 124]{pagonis_strange_1999})
\end{quote}
One can construct states where, at a moment, $\phi$ is only non-zero within a given region and then, very soon after, $\phi$ is non-zero far away (at spacetime points space-like separated from the original region).  At first glance, that might look like superluminal propagation.  But, by our earlier second-order standard for locality, it is not.  If $\partial \phi / \partial t$ is non-zero outside the region then there are important things happening out there, even if $\phi$ itself happens to be momentarily zero.  When you use the Klein-Gordon equation (which is second-order) to determine the evolution of $\phi$ somewhere far from the specified region, that evolution will take as input both $\phi$ (which is zero) and $\partial \phi / \partial t$ (which is not zero).

This dissolution of the superluminal propagation problem relies on the common understanding of the Klein-Gordon dynamics as second-order, taking as input both $\phi$ and its time derivative, $\partial \phi / \partial t$.  By contrast, if the dynamics were first-order then perhaps we should be troubled by cases where $\phi$ is initially only non-zero in some region and soon after is non-zero at far space-like separated points.  \textcite{pagonis_strange_1999} and \textcite[pg.\ 359]{tumulka_foundations_2022} explicitly argue that the Klein-Gordon dynamics should be viewed as first-order. 
\textcite[pg.\ 123--124]{pagonis_strange_1999}  write that the first-order equation\footnote{There are other first-order reformulations of the Klein-Gordon dynamics: \textcite[sec.\ 1.6 and 1.8]{greiner_relativistic_2000} gives a first-order Schr\"{o}dinger form of the Klein-Gordon equation with a two-component wave function and modifies that to arrive at the first-order Feshbach-Villars representation.}
\begin{equation}
i \hbar \frac{\partial}{\partial t}\phi = \sqrt{m^2 c^4 - \hbar^2 c^2 \nabla^2}\phi
\label{firstorderKG}
\end{equation}
\begin{quote}
``supercedes the KG [Klein-Gordon] equation, because there is a one-to-one correspondence between positive energy solutions of the Klein-Gordon equation
and specifications of (smooth) $\phi$ on a spacelike hyperplane and [\eqref{firstorderKG}] implies the KG equation by iteration.''
\end{quote}
In favor of \eqref{firstorderKG}, \textcite[pg.\ 359]{tumulka_foundations_2022} writes that
\begin{quote}
``the second-order Klein-Gordon equation includes contributions of negative energy and is therefore widely avoided.''
\end{quote}
The operator under the square root in \eqref{firstorderKG} looks a bit odd, but can be rendered precise by Fourier transforming $\phi$, writing the operator in terms of the wave vector $\vec{k}$, and Fourier transforming back (as in \cite[pg.\ 56]{schweber_introduction_1961}):
\begin{equation}
i \hbar \frac{\partial}{\partial t}\phi (\vec{x}\,) =\frac{1}{(2\pi)^3} \int d^3 k \; d^3\vec{y}\ e^{i \vec{k}\cdot(\vec{x}-\vec{y}\,)}\sqrt{m^2 c^4 + \hbar^2 c^2 k^2}\phi(\vec{y}\,)
\ .
\label{firstorderKG2}
\end{equation}
This first-order equation for the evolution of $\phi$ is clearly non-local, but Fleming and Butterfield are correct that we can use the equation to pick out the allowed free evolutions for positive energy states.  We do not see a compelling argument in the above quotations from Fleming, Butterfield, and Tumulka as to why the first-order equation should be viewed as more fundamental.  In response to Fleming and Butterfield: It is true that one can derive the second-order equation from the first-order, but you can also go the other way and use the second-order equation to show that states formed from positive energy modes will satisfy the first-order equation.  In response to Tumulka: Using the second-order equation only fixes the dynamics and we still have the freedom to choose whether to allow physical states that include negative energy modes (depending on their ultimate utility for the representation of either particles or antiparticles).

\subsection{The Dirac Equation}\label{DEsection}

In this section, we prove the locality of the Dirac equation.\footnote{Locality is proved for the standard Dirac equation.  If the alternative Foldy-Wouthuysen representation is taken as fundamental, the dynamics are non-local (\cite[pg.\ 28]{thaller_dirac_1992}).}  While this property of the Dirac equation is known, we have not seen a proof that neatly parallels the proofs of locality for the Maxwell and Klein-Gordon equations presented earlier.\footnote{\textcite[sec.\ 1.5]{thaller_dirac_1992} gives a very different proof of the same result.} Here we present such a proof.

Let us begin with the free Dirac equation:
\begin{equation}
\bigg( i \hbar \frac{\partial}{\partial t} +i \hbar c\, \gamma^0\vec{\gamma}\cdot\vec{\nabla} - \gamma^0 m c^2\bigg)\psi = 0
\ .
\label{dirac}
\end{equation}
Because this is a first-order equation, the standard for locality (from the end of section \ref{EMsection}) is whether two solutions $\psi_1$ and $\psi_2$ that agree within some sphere $R$ at $t=0$ will agree within the contracting light-cone with $R$ as its base (figure \ref{frustumfig})---with no requirement that the first time-derivatives of $\psi_1$ and $\psi_2$ agree on $R$.  Let us define $\psi_d$ to be the difference between $\psi_1$ and $\psi_2$: $\psi_d=\psi_1-\psi_2$.  Because the free Dirac equation is linear, $\psi_d$ will obey the free Dirac equation if $\psi_1$ and $\psi_2$ do.

From the Dirac equation, one can prove a conservation law that might be regarded as describing the local conservation of probability or charge (depending on whether the Dirac equation is being viewed as part of relativistic quantum mechanics or classical field theory).  The difference $\psi_d$ must thus obey:
\begin{equation}
\frac{\partial}{\partial t}|\psi|^2+\vec{\nabla}\cdot\bigg(c \psi^\dagger \gamma^{0} \vec{\gamma} \psi \bigg)=0
\ .
\label{diraccontinuity}
\end{equation}
This continuity equation is of the right form to apply the four-dimensional divergence theorem, as in \eqref{phidwave3}.  Let us integrate \eqref{diraccontinuity} over the interior of the frustum $F$ and use the four-dimensional divergence theorem to rewrite this four-dimensional volume integral as an integral over the three-dimensional ``surface'' comprised of the base $R$, the top $R^-$, and the edge $E$,
\begin{align}
0&=\int_{R\cup R^- \cup E} d^3\vec{x} \bigg(n_t |\psi_d|^2 -\vec{n}\cdot\bigg(c \psi_d^\dagger \gamma^{0} \vec{\gamma} \psi_d \bigg)\bigg)
\nonumber
\\
&=-\int_{R} d^3\vec{x} |\psi_d|^2 +\int_{R^-} d^3\vec{x} |\psi_d|^2
\nonumber
\\
&\quad\quad\quad\quad +\int_{E} d^3\vec{x} \left[n_t |\psi_d|^2 - \vec{n}\cdot\bigg(c \psi_d^\dagger \gamma^{0} \vec{\gamma} \psi_d\bigg)\right]
\ .
\label{diracint}
\end{align}
The integral over $R$ vanishes because, by supposition, $\psi_1=\psi_2$ on $R$ and thus $\psi_d$ is zero on $R$.  We can insert our earlier expression for the four-dimensional unit vector \eqref{4Dnormal} normal to the edge $E$ of the frustum to express the integral over $E$ as
\begin{align}
&\frac{c}{\sqrt{c^2+1}} \int_{E} d^3\vec{x} \left[|\psi_d|^2 - \frac{\vec{x} - \vec{x}_f}{c|\vec{x} - \vec{x}_f|} \cdot\bigg(c \psi_d^\dagger \gamma^{0} \vec{\gamma} \psi_d\bigg)\right]
\end{align}
This can be rewritten as a sum of squares
\begin{align}
&\frac{c}{\sqrt{c^2+1}} \int_{E} d^3\vec{x} \left[\left|\psi_d - \frac{1}{2}\frac{\vec{x} - \vec{x}_f}{|\vec{x} - \vec{x}_f|} \cdot\bigg(\gamma^{0} \vec{\gamma} \psi_d\bigg)\right|^2 - \frac{1}{4}\left|\frac{\vec{x} - \vec{x}_f}{|\vec{x} - \vec{x}_f|} \cdot\bigg(\gamma^{0} \vec{\gamma} \psi_d\bigg)\right|^2 \right]
\ ,
\end{align}
using the properties of the Dirac gamma matrices that (a) the zero matrix is Hermitian: $\gamma^{0\dagger}=\gamma^{0}$, (b) the $x$, $y$, and $z$ matrices are anti-Hermitian: $\gamma^{i\dagger}=-\gamma^{i}$, and (c) the matrices obey the anticommutation relations $\{\gamma^\mu,\gamma^\nu\}=2\eta^{\mu\nu}$ with metric signature (+ - - -).  Employing the same tools again, the final term becomes
\begin{align}
&\frac{c}{\sqrt{c^2+1}} \int_{E} d^3\vec{x} \left[\left|\psi_d - \frac{1}{2}\frac{\vec{x} - \vec{x}_f}{|\vec{x} - \vec{x}_f|} \cdot\bigg(\gamma^{0} \vec{\gamma} \psi_d\bigg)\right|^2 - \frac{1}{4}\psi_d^\dagger \bigg(\frac{\vec{x} - \vec{x}_f}{|\vec{x} - \vec{x}_f|} \cdot\vec{\gamma} \bigg)\bigg(\frac{\vec{x} - \vec{x}_f}{|\vec{x} - \vec{x}_f|} \cdot\vec{\gamma} \bigg)\psi_d \right]
\ .
\end{align}
The unit three-vector $\frac{\vec{x} - \vec{x}_f}{|\vec{x} - \vec{x}_f|}$ picks out a direction and taking the dot product with $\vec{\gamma}$ gives us a gamma matrix associated with that direction.  Like $\gamma^1$, $\gamma^2$, or $\gamma^3$, $\frac{\vec{x} - \vec{x}_f}{|\vec{x} - \vec{x}_f|} \cdot\vec{\gamma}$ will square to $-1$.  Thus, our integral over $E$ becomes
\begin{equation}
\frac{c}{\sqrt{c^2+1}} \int_{E} d^3\vec{x} \left[\left|\psi_d - \frac{1}{2}\frac{\vec{x} - \vec{x}_f}{|\vec{x} - \vec{x}_f|} \cdot\bigg(\gamma^{0} \vec{\gamma} \psi_d\bigg)\right|^2 + \frac{1}{4}|\psi_d|^2 \right]
\ .
\end{equation}
This is a sum of squares and thus must be $\geq 0$.  Returning to \eqref{diracint}, this means that the integral over $R^-$ must satisfy
\begin{equation}
\int_{R^-} d^3\vec{x} |\psi_d|^2 \leq 0
\end{equation}
The only way that this inequality can hold is if $|\psi_d|^2=0$ (and thus $\psi_d=0$) throughout $R^-$. The two solutions $\psi_1$ and $\psi_2$ that agreed on $R$ must agree on $R^-$.  This is true regardless of where in time we cut the slice $R^-$ along the contracting light-cone, and thus $\psi_1$ and $\psi_2$ must agree throughout the contracting light-cone.  This completes our proof that the Dirac equation satisfies the same standard of relativistic locality as electromagnetism and the Klein-Gordon equation.

\section{Quantum Field Theory}\label{QFTsection}

Having seen that a single standard for locality holds for classical electromagnetism, the Klein-Gordon, and the Dirac equation, we next embark on the task of showing that the same standard holds for QFT.  Because our most fundamental quantum theories are cast in the framework of QFT, we believe that establishing the locality of QFT (without collapse) suffices to show that the many-worlds interpretation is local.  In this section, we will follow the standard convention for QFT and set $\hbar=c=1$.

To apply the standard for relativistic locality from the end of section \ref{EMsection}, we first need a way of assigning states to regions at a time in QFT.  \textcite{wallace_quantum_2010} and \textcite[ch.\ 8]{wallace_emergent_2012} have considered this question in the context of the many-worlds interpretation and proposed a way of assigning states to regions of spacetime and to regions of space at a time.  Wallace and Timpson also claim that their spacetime states evolve locally.  Although we broadly agree with their approach and see ours as compatible with theirs, we think that there are many details left to be filled in by their quick treatment. 

Wallace and Timpson propose that we should assign a ``density operator'' or ``reduced density matrix'' to a region of space at a time to give the state of that region.  That much we agree on.  However, when presenting the nature of this density operator in relativistic QFT, they seek to avoid adopting a preferred basis\footnote{The approach that we prefer for assigning states to regions (section \ref{WFFSsection}) makes use of a particular basis (in our example, configurations of the field $\phi$).  Still, one need not take this to be a metaphysically preferred basis. One could maintain that the states of regions can be expressed equally well in other bases.  This would give a way of responding to the following concern that \textcite[pg.\ 707--708]{wallace_quantum_2010} raise: ``there is no single preferred choice of fields by which a QFT can be specified.  A number of results from QFT \dots suggest that a single QFT can be equivalently described in terms of several different choices
of field observable, with nothing in particular to choose between them.''} (such as particle or field configurations) and end up taking an algebraic approach that we see as overly abstract and general (see \cite{wallace_quantum_2010}; \cite[pg.\ 301]{wallace_emergent_2012}; \cite{swanson_how_2020}).  In this section, we explore two competing proposals as to how one might assign density matrix states to regions that we take to be more ontologically precise.  The first option, which we prefer, begins from the field wave functional approach to QFT and leads to locality. The second option begins from the competing particle Fock space approach and either fails in assigning states to regions or leads to non-locality.

\textcite[pg.\ 302--303]{wallace_emergent_2012} appeals to a standard result from QFT to argue that the states that he and Timpson identify will evolve locally:
\begin{quote}
    ``In a quantum field theory, the quantum state of any region depends only on the quantum state of some cross-section of the past lightcone of that region. Disturbances cannot propagate into that lightcone. This follows from the well-known fact that spacelike separated field operators commute: any disturbance outside the past lightcone of a region $R$ can be represented on the global quantum state as the action on that state of a unitary operator built out of some operators localized outside the past light cone of $R$, and since those operators commute with all operators localized in $R$, they have no effect on the expectation values of operators in $R$, and so no effect on the physical state of $R$.''
\end{quote}
While we agree with the verdict in the first sentence and will make use of the same result about the commutation of space-like separated operators to prove it, we do not see this quick argument as sufficient.  As will become clear in what follows, it takes a significant amount of work to settle the question of fundamental locality for a particular concrete choice of ontology (applying the standard for locality from section \ref{EMsection}).

In what follows, we do not adopt a particular strategy for handling the UV divergences that arise in QFT, but assume that they can be handled somehow.  Our proof of locality for a field approach to QFT leaves some gaps to be filled in and certainly not up to the standards of mathematical rigor found in algebraic approaches to quantum field theory.\footnote{We seek to operate at the level of rigor of a typical QFT textbook. Problems may arise at higher levels of rigor. For instance, the way that we assign density operators to regions by taking partial traces \eqref{densitymatrixoperators} might run into technical difficulties having to do with physics at the UV scale and the question of whether the Hilbert space is separable (see \cite{swanson_how_2020}).  Setting these issues aside and working at a lower level of rigor, it is already non-trivial to prove the relativistic locality of unitary quantum field theory in terms of state evolution.} Still, we think it suffices to illustrate the locality of QFT.

\subsection{Wave Functionals and Field States for Regions}\label{WFFSsection}

On what can be called a ``field approach'' to QFT, the universal quantum state is given by a wave functional $\Psi$ that assigns amplitudes to different possible classical field configurations (different specifications of the field values at every point in space).\footnote{The field approach is presented in \textcite{jackiw_schrodinger_1987}; \textcite{hatfield_quantum_1992}; \textcite[ch.\ 11]{bohm_undivided_1993}; \textcite{struyve_pilot-wave_2010}; \textcite{sebens_fundamentality_2022}.}  This is analogous to the way that a particle wave function assigns amplitudes to different possible classical particle configurations (different specifications of the locations of point particles).  Adopting the Schr\"{o}dinger picture, the wave functional evolves by a Schr\"{o}dinger equation that takes the usual form,
\begin{equation}
i \frac{d}{d t}\Psi = \hat{H}\Psi
\ .
\label{SE}
\end{equation}

When doing non-relativistic quantum mechanics for a set of non-identical particles without spin, you can express the state of a subset of the particles as a reduced density matrix by taking the universal density matrix and tracing over the possible locations of the particles outside the subset (\cite[sec.\ 3]{durr_role_2005}).  In a field approach to QFT, you can express the state of a region of space by taking the universal density matrix and tracing over the field values outside that region.  The universal wave functional for a single bosonic scalar field $\phi$ can be written as
\begin{equation}
\Psi\left[\phi\right]=\Psi\left[\phi^R;\phi^{\bar{R}}\right]
\ ,
\end{equation}
where $\phi$ is understood to be a function that ranges over the entirety of space, $\phi^R$ only assigns field values to points within $R$, and $\phi^{\bar{R}}$ only assigns field values to points within $\bar{R}$. The universal density matrix (which we have assumed to be pure) can be written in terms of the universal wave functional as
\begin{align}
\rho[\phi;\phi'] &= \Psi\left[\phi\right]\Psi^*\left[\phi'\right]=\Psi\left[\phi^R;\phi^{\bar{R}}\right]\Psi^*\left[\phi'^R;\phi'^{\bar{R}}\right]
\ ,
\end{align}
Tracing over the field values outside of $R$, we can write the reduced density matrix for a region $R$ as
\begin{align}
\rho^R[\phi^R;\phi'^R] &= \int \mathcal{D}\phi^{\bar{R}}\ \Psi\left[\phi^R;\phi^{\bar{R}}\right]\Psi^*\left[\phi'^R;\phi^{\bar{R}}\right]
\ ,
\label{WFstateofA}
\end{align}
where $\int \mathcal{D}\phi^{\bar{R}}$ denotes integrals over the field values $\phi^{\bar{R}}(\vec{x})$ at each point $\vec{x}$ within $\bar{R}$.\footnote{See \textcite[sec.\ 9.3]{hatfield_quantum_1992} for the $\mathcal{D}\phi$ integral notation.}  Thus, we have arrived at precise states for regions in QFT.  (Such states are described in, e.g., \cite[pg.\ 445]{holzhey_geometric_1994}; \cite[eq.\ 3.4.28]{susskind_introduction_2005}.)

We can move from the above expressions for the universal and reduced density matrices as functions of two field configurations to operator expressions\footnote{\textcite[sec.\ 3]{durr_role_2005} present similar operator and function versions of the reduced density matrices for collections of particles in non-relativistic quantum mechanics.} (which will be of use later) by inserting integrals over possible field configurations and field eigenstates,
\begin{align}
\hat{\rho} &= \int \mathcal{D}\phi\mathcal{D}\phi'\ \rho[\phi;\phi'] |\phi\rangle\langle\phi'|
\nonumber
\\
\hat{\rho}^R &= \int \mathcal{D}\phi^R\mathcal{D}\phi'^R\ \rho^R[\phi^R;\phi'^R] |\phi^R\rangle\langle\phi'^R|
\ .
\label{densitymatrixoperators}
\end{align}
The field eigenstates $|\phi\rangle$ are such that a field operator $\hat{\phi}(\vec{x})$ acting on $|\phi\rangle$ returns the value of $\phi$ at $\vec{x}$.\footnote{See \textcite[eq.\ 10.5]{hatfield_quantum_1992}.}  The field eigenstates $|\phi^R\rangle$ within $R$ return the values of $\phi^R$ when acted upon by a field operator within $R$.  Note that the eigenstates $|\phi^R\rangle$ are partial states, only describing the field degrees of freedom within $R$.  To get a full eigenstate for the field everywhere, you would need to combine this with a field configuration $\phi^{\bar{R}}$ in the complement of $R$ and write the state as $|\phi^{\bar{R}},\phi^{R}\rangle$ or $|\phi^{\bar{R}}\rangle|\phi^{R}\rangle$.

Here we have focused on finding reduced density matrix states for regions starting from a wave functional for a single bosonic scalar field.  The extension to other bosonic fields (or multiple bosonic fields) is straightforward.  Formally, one could use the same method to find reduced density matrix states for regions starting from a wave functional for a single fermionic field.  However, puzzles arise relating to the use of Grassmann numbers in such wave functionals.  (The field approach to QFT faces serious challenges regarding Grassmann numbers that must be addressed for the approach to be viable---see \cite[sec.\ 5.1]{sebens_fundamentality_2022}.)

\subsection{Wave Functional Dynamics and Relativistic Locality}\label{WFDsection}

To show that QFT meets the standard for relativistic locality from the end of section \ref{EMsection}, we must show that specifying the state of some sphere $R$ at $t=0$ fixes the future states within the contracting light-cone of $R$.  Because the dynamics \eqref{SE} are first-order, the state of $R$ alone should be sufficient to determine that evolution. We do not need to specify any time derivatives.

Let us focus first on the simplest case of a bosonic QFT for a single free real scalar field (the Klein-Gordon field).  Adopting the Schr\"odinger picture (as is rarely done but perfectly permissible in QFT), the dynamics of the universal density matrix \eqref{densitymatrixoperators} are given by the von Neumann equation,
\begin{equation}
i \frac{d}{d t}\hat{\rho} = [\hat{H},\hat{\rho}]
\ ,
\label{vonneumann}
\end{equation}
which (for a time-independent Hamiltonian) has the solutions
\begin{equation}
\hat{\rho}(t) = e^{-i \hat{H} t}\hat{\rho}(0)e^{i \hat{H} t}
\ .
\label{vonneumannsol}
\end{equation}
For a free real scalar quantum field, we can take the Hamiltonian $\hat{H}$ to be the integral over all space of the Hamiltonian density operator
\begin{align}
\hat{\mathscr{H}}(\vec{x})&=\frac{1}{2}\bigg(\hat{\pi}^2(\vec{x})+|\vec{\nabla}\hat{\phi}(\vec{x})|^2+m^2\hat{\phi}(\vec{x})^2\bigg)
\ ,
\label{scalarhamiltonian}
\end{align}
where the $\hat{\pi}$ operator (called the ``conjugate field momentum'') that appears here can be understood as a functional derivative: $\hat{\pi}(\vec{x})=-i\frac{\delta}{\delta \phi(\vec{x})}$ (\cite[eq.\ 10.9]{hatfield_quantum_1992}).  (The common expression for the Hamiltonian above should arguably be normal-ordered, but let us set that complication aside here.\footnote{See \textcite[pg.\ 45--46]{hatfield_quantum_1992}; \textcite[pg.\ 81]{greiner_field_1996}.})  Although at this point it makes things more concrete to have a particular Hamiltonian in mind, the proof will end up applying more generally across different Hamiltonians.

The key property of the above free Hamiltonian, for the purposes of proving locality, is that when it is used to construct Heisenberg picture operators, local operators will commute at space-like separation,
\begin{equation}
[\hat{\mathcal{O}}_1(\vec{x},t),\hat{\mathcal{O}}_2(\vec{x}',t')]= 0 \quad \mbox{ if $(\vec{x},t)$ and $(\vec{x}',t')$ are space-like separated.}
\label{gencausalitycond}
\end{equation}
This commutation property is standardly taken to be central in textbook proofs of ``causality'' (or ``local causality'' or ``microcausality'') and it is among the axioms in axiomatic approaches to relativistic QFT.\footnote{See \textcite{earman_relativistic_2014}; \textcite{calderon_causal_2024}.}  We will not use a particular Hamiltonian, like \eqref{scalarhamiltonian}, to prove that \eqref{gencausalitycond} holds for certain local operators here, but instead simply show how this commutation property (once established) can be used to prove locality.  Although \eqref{gencausalitycond} may sometimes be presented as a locality condition, we will see that it takes some work to go from this commutation property to a demonstration that QFT satisfies the standard for relativistic locality from section \ref{EMsection}.    From \eqref{gencausalitycond}, with $t'=0$, it follows that
\begin{align}
[e^{-i \hat{H} t}\hat{\mathcal{O}}_2(\vec{x}')e^{i \hat{H} t},\hat{\mathcal{O}}_1(\vec{x})]&= 0
\quad \mbox{ if $(\vec{x},t)$ and $(\vec{x}',0)$ are space-like separated,}
\label{gencausalitycond2}
\end{align}
which is a form that will be useful later.

As a special case of \eqref{gencausalitycond}, the Heisenberg picture field operators commute at space-like separation,\footnote{See, e.g., \textcite[sec.\ 2.4]{peskin_introduction_1995}; \textcite[sec.\ 4.4]{greiner_field_1996}; \textcite[pg.\ 46]{hatfield_quantum_1992}; \textcite[sec.\ 2.6.1]{tong_lectures_2007}.}
\begin{equation}
[\hat{\phi}(\vec{x},t),\hat{\phi}(\vec{x}',t')]= 0
\quad \mbox{ if $(\vec{x},t)$ and $(\vec{x}',t')$ are space-like separated.}
\label{causalitycond}
\end{equation}
In their discussion of \eqref{causalitycond}, Peskin and Schroeder first show that $\langle \Omega| \hat{\phi}(\vec{x},t)\hat{\phi}(\vec{x}',t') |\Omega\rangle \neq 0$ (where $|\Omega\rangle$ is the vacuum) when $(\vec{x},t)$ and $(\vec{x}',t')$ are space-like separated, which they interpret as showing that a particle \emph{can} propagate from $(\vec{x},t)$ to $(\vec{x}',t')$---a faster-than-light motion. That seems troubling, but Peskin and Schroeder reassure the reader:
\begin{quote}
``To really discuss causality, however, we should ask not whether particles can propagate over spacelike intervals, but whether a \emph{measurement} performed at one point can affect a measurement at another point whose separation from the first is spacelike.''
\end{quote}
The commutation of field operators at space-like separation \eqref{causalitycond} is then taken to show that such measurements will be unable to affect one another.

Peskin and Schroeder's focus on measurements is standard in QFT textbooks,\footnote{See \textcite[pg.\ 222--223]{schweber_introduction_1961}; \textcite[pg.\ 38]{hatfield_quantum_1992}; \textcite[pg.\ 198]{weinberg_quantum_2005}; \textcite[pg.\ 219]{schwartz_quantum_2014}; \textcite[pg.\ 102--103]{greiner_field_1996}; \textcite[sec.\ 2.6--2.7]{tong_lectures_2007}.} but we do not need to be deterred by Peskin and Schroeder's warning.  To ``really discuss causality,'' we should ask what is happening in nature, whether or not measurements are being conducted.  In an Everettian (many-worlds) approach to QFT, we can ask such questions and we will see that there is no superluminal propagation.  The commutation of operators at space-like separation will be central to that proof, but there are further steps that must be traversed to go from \eqref{gencausalitycond} to a proof of locality that resembles our earlier proofs for simpler theories. 

As with our earlier proofs of locality for the Maxwell, Klein-Gordon, and Dirac equations, let us begin with two distinct evolving density matrices that are solutions to the von Neumann equation \eqref{vonneumann}, $\hat{\rho}_1(t)$ and $\hat{\rho}_2(t)$, and agree on the state within $R$ at $t=0$.  One can introduce a matrix $\hat{\rho}_d(t)=\hat{\rho}_1(t)-\hat{\rho}_2(t)$ that is the difference between these solutions.  This matrix is a solution to the von Neumann equation but is not a \emph{density matrix} because it has trace zero: 
\begin{align}
\mbox{tr}\,\hat{\rho}_d(t)&=\mbox{tr}\,\hat{\rho}_1(t)-\mbox{tr}\,\hat{\rho}_2(t)=1-1=0
\ ,
\label{rhodtrace}
\end{align}
where $\mbox{tr}\,\hat{\rho}_1(t)=\mbox{tr}\,\hat{\rho}_2(t)=1$ as these are properly normalized states.  Noting that it is not a density matrix, let us call $\hat{\rho}_d(t)$ a ``difference matrix''.

By the assumption that $\hat{\rho}_1(t)$ and $\hat{\rho}_2(t)$ agree on the state within $R$ at $t=0$, the difference matrix $\hat{\rho}_d(t)$ will yield zero as the reduced difference matrix for $R$ at $t=0$ (integrating over the possible field configurations for $\bar{R}$):
\begin{align}\label{difference}
\hat{\rho}_d^{R}(0)&=\mbox{tr}_{\bar{R}}\hat{\rho}_d(0)
\nonumber
\\
&=\int \mathcal{D}\phi^{\bar{R}}\  \langle \phi^{\bar{R}} | \hat{\rho}_d(0) | \phi^{\bar{R}} \rangle
\nonumber
\\
&=0
\ .
\end{align}
The task at hand, then, is to show that at an arbitrary later time $t$, $\hat{\rho}_d^{R^-}(t)=0$, or in other words, that $\hat{\rho}_1(t)$ and $\hat{\rho}_2(t)$ agree on the state within the slice $R^-$ of the contracting light-cone that has $R$ as its base (recall figure \ref{lightcones}).

Because $\hat{\rho}_1(0)$ and $\hat{\rho}_2(0)$ agree within $R$, the universal difference matrix $\hat{\rho}_d(0)$ initially assigns no reduced difference matrix to $R$ and $\hat{\rho}_d(0)$ can be written as a local operator restricted to $\bar{R}$ (as in \eqref{densitymatrixoperators}),
\begin{align}\label{localoperatorrho}
\hat{\rho}_d(0) &=\hat{\rho}^{\bar{R}}_d(0)= \int \mathcal{D}\phi^{\bar{R}}\mathcal{D}\phi'^{\bar{R}}\ \rho_d^{\bar{R}}[\phi^{\bar{R}};\phi'^{\bar{R}};0] |\phi^{\bar{R}}\rangle\langle\phi'^{\bar{R}}|
\ ,
\end{align}
where $\rho_d^{\bar{R}}[\phi^{\bar{R}};\phi'^{\bar{R}};0]$ are just complex numbers associated with each pair of possible field configurations $\phi^{\bar{R}}$ and $\phi'^{\bar{R}}$ on the complement of $R$ at $t=0$.  The next step is to show that $\hat{\rho}_d(t)= e^{-i \hat{H} t}\hat{\rho}_d(0)e^{i \hat{H} t}$ is a local operator restricted to $\bar{R}^-$.  To do so, it should be sufficient to show that it commutes with any operator in $R^{-}$:
\begin{equation}
[ e^{-i \hat{H} t}\hat{\rho}_d(0)e^{i \hat{H} t},\hat{\mathcal{O}}(\vec{x})] = 0 \quad \mbox{for any $\hat{\mathcal{O}}(\vec{x})$ with $\vec{x} \in R^{-}$.}
\label{finalcommutation}
\end{equation}
Here we can appeal to the commutation at space-like separation that is standardly taken to establish ``causality'' in QFT, \eqref{gencausalitycond} in the form of \eqref{gencausalitycond2}.  Assuming that $\hat{\rho}_d(0)$ can be expanded in terms of local operators at points within $\bar{R}$ (which should be possible given \eqref{localoperatorrho}), \eqref{gencausalitycond2} ensures that each such operator will commute with $\hat{\mathcal{O}}(\vec{x})$ and thus that \eqref{finalcommutation} holds.  Having established that $\hat{\rho}_d(t)$ is a local operator restricted to $\bar{R}^-$, tracing over $\bar{R}^-$ will yield zero:
\begin{align}
\hat{\rho}_d^{R^-}(t)&=\mbox{tr}_{\bar{R}^-}\hat{\rho}_d(t)=\int \mathcal{D}\phi^{\bar{R}^-}\  \langle \phi^{\bar{R}^-} |e^{-i \hat{H} t}\hat{\rho}_d(0)e^{i \hat{H} t}|\phi^{\bar{R}^-} \rangle
\\
\nonumber
&=\int \mathcal{D}\phi^{\bar{R}^-}\mathcal{D}\phi^{R^{-}}\  \langle \phi^{\bar{R}^-}|\langle\phi^{R^{-}} |e^{-i \hat{H} t}\hat{\rho}_d(0)e^{i \hat{H} t}|\phi^{R^{-}} \rangle|\phi^{\bar{R}^-}\rangle=\mbox{tr}\hat{\rho}_d(t)=0
\ ,
\end{align}
where the second line inserts $\int \mathcal{D}\phi^{R^{-}}\  \langle\phi^{R^{-}} |\phi^{R^{-}} \rangle=1$ and uses \eqref{rhodtrace}.\footnote{The equation $\int \mathcal{D}\phi^{R^{-}}\  \langle\phi^{R^{-}} |\phi^{R^{-}} \rangle=1$ is similar to how in non-relativistic quantum mechanics the integral of the inner product of a position eigenstate with itself is 1: $\int d^3\vec{x}\  \langle \vec{x} | \vec{x} \rangle=1$.}  The fact that $\hat{\rho}_d^{R^-}(t)=0$ means that $\hat{\rho}^{R^-}_1(t)$ and $\hat{\rho}^{R^-}_2(t)$ must agree and our proof is finished. (The two recent uses of ``should be,'' before and after \eqref{finalcommutation}, leave gaps that would need to be filled in to arrive at a more thorough proof.)

This proof can be extended immediately to interacting theories because the commutation of local operators at space-like separation \eqref{gencausalitycond} is a result that holds across bosonic relativistic quantum field theories, not just for the free Hamiltonian \eqref{scalarhamiltonian}.  Indeed, that commutation condition is a standard that must be met for a bosonic quantum field theory to be classified as a ``local quantum field theory'' (see \cite[pg.\ 37]{tong_lectures_2007}).\footnote{\textcite[sec.\ 6]{williams_cluster_2024} formulate Weinberg's principle of ``cluster decomposition'' as a constraint on acceptable Hamiltonians that might suffice for guaranteeing the necessary commutation relations hold.}

For fermionic fields, a common textbook line is that even though the Heisenberg picture field operators satisfy \emph{anticommutation} relations instead of commutation relations, there is still no violation of causality (locality) because pairs of field operators (``bilinears'') commute at space-like separation and \emph{observables} (the kinds of operators that you can actually measure) will be pairs of field operators (see \cite[sec.\ 5.4]{tong_lectures_2007}; \cite[pg.\ 219]{schwartz_quantum_2014}; \cite[exercise 5.5]{greiner_field_1996}).  That discussion of measurement is insufficient for our purposes, as we would like to prove that relativistic locality holds for the time evolution of quantum fields regardless of whether any measurements are being conducted.  Still, we can potentially use the standard results about anticommutation at space-like separation to prove locality for fermionic fields.  If it can be proven that, for a fermionic field, the initial difference matrix $\hat{\rho}_d(0)=\hat{\rho}^{\bar{R}}_d(0)$ is a sum of pairs of anticommuting local operators, then that will commute with operators at space-like separation as in \eqref{finalcommutation} and locality can be proven as above.

\subsection{Interlude: Local Dynamics, Non-Separable States}

We have now seen that QFT, as a theory of fields, can be shown to meet the same standard for relativistic locality as electromagnetism, the Klein-Gordon equation, the Dirac equation. In this sense, QFT is relativistically local and there is no influence from outside the past light-cone on any spatial region.

However, one might be concerned that our construction of the reduced density matrix giving the state of a region $R$ requires starting with the full universal state and tracing over the region \textit{outside} of $R$. One could argue that the reduced density matrix for $R$ is not local to $R$ because it depends by its definition on what is happening outside of $R$. 

To address this concern, we want to emphasize the distinction between \emph{non-separability} and \emph{relativistic non-locality}. Following \textcite{wallace_emergent_2012}, we can define non-separability as such:\footnote{See \textcite{healey_holism_1991}, \textcite{wallace_quantum_2010}, \textcite{myrvold_lessons_2015}, and \textcite[sec.\ 6]{williams_cluster_2024} for similar discussions of non-separability.} 
\begin{quote}
    A theory is non-separable if, given two regions $A$ and $B$, a complete specification of the states of $A$ and $B$ separately fails to fix the state of the combined system $A + B$. That is, there are additional facts -- nonlocal facts, if we take $A$ and $B$ to be spatially separated -- about the combined system, in addition to the facts about the two individual systems. (\cite[pg. 293]{wallace_emergent_2012})
\end{quote}
As an example, consider an idealized setup where there is a single spin-$1/2$ particle in region $A$, another in region $B$, and we are only concerned with representing facts about the spins of each particle.  The total wave function might be in the spin singlet state where the spins are opposite,
\begin{equation}
    |\Psi\rangle = \frac{1}{\sqrt{2}} \bigg(|\uparrow_z\rangle_A|\downarrow_z\rangle_B - |\downarrow_z\rangle_A|\uparrow_z\rangle_B \bigg)
    \label{singlet}
\end{equation}
or the triplet state
\begin{equation}\label{triplet}
    |\Psi\rangle = \frac{1}{\sqrt{2}} \bigg(|\uparrow_z\rangle_A|\downarrow_z\rangle_B + |\downarrow_z\rangle_A|\uparrow_z\rangle_B \bigg)
    \ .
\end{equation}
These are distinct total states (that make different predictions as to whether the $x$ spins of the two particles could be aligned when measured) that yield the same reduced density matrices for $A$ and $B$.  Similarly, in QFT the reduced density matrix states of two regions $A$ and $B$ will not fix the state of $A \cup B$ because we will be missing facts about the entanglement between these regions.

Non-separability is not ruled out by special relativity.  Special relativity only forbids violations of relativistic locality (as defined in section \ref{EMsection}). Relativistic locality is a property of the dynamics, prohibiting instantaneous action at a distance or any other faster-than-light interactions. Newtonian gravity violates this standard, though it has a separable ontology.  QFT, by contrast, has a non-separable ontology but satisfies the standard of relativistic locality.

One might worry that non-separability could lead to violations of relativistic causality, because the non-separable states of widely spread-out composite systems can immediately change when acted upon at one location.\footnote{We would like to thank Jacob Barandes for discussion of this point.}  As an example, consider taking the spin singlet state in \eqref{singlet} and applying a unitary transformation on the particle in $A$ alone (e.g., via the application of an appropriate magnetic field) which flips its $z$-spin, yielding:
\begin{equation}\label{singletflipped}
    |\Psi\rangle = \frac{1}{\sqrt{2}} \bigg(|\downarrow_z\rangle_A|\downarrow_z\rangle_B - |\uparrow_z\rangle_A|\uparrow_z\rangle_B \bigg)
    \ .
\end{equation}
That is, via a unitary transformation acting locally on the particle in $A$, the entanglement relation of perfect anti-correlation between the $z$-spins of the two particles is transformed into an entanglement relation of perfect correlation. A local action within $A$ has changed the global state of the two particles, a state that extends into a region, $B$, that is space-like separated from $A$. Furthermore, note that the reduced density matrices of the two particles are the same in both \eqref{singlet} and \eqref{singletflipped}. This local action leaves the local states of the $A$-particle and $B$-particle unchanged while causing global changes in the state of the two particles together.

Although this might appear to be a violation of relativistic locality, it is not.  The global state of the two particles, located in region $A \cup B$, has changed due to a cause within region $A$, a part of $A \cup B$.  The cause is local to the effect. The cause happens at a place where the global state is, rather than at any spatial separation from the global state.  An analogy: a country's global state of having a certain population changes when a birth occurs on one coast, but this does not involve any action at a distance or superluminal interaction with the opposite coast.  Granted, the situation with quantum entanglement is not entirely analogous. The birth changes the local state of the town on the coast,  whereas the transformation of the particle in $A$ leaves its local state unchanged (while still changing the global state of the two particles). However, because in both cases the cause is local to the effect, neither presents a problem for relativistic locality.

\subsection{Fock Space and Particle States for Regions}\label{FSPsection}

While we think that QFT and its local states are best discussed in terms of wave functionals (interpreting QFT as a theory of fields),\footnote{See \textcite{sebens_fundamentality_2022} for a defense of the field approach and an acknowledgment of the challenges facing the approach (including the use of Grassmann numbers when dealing with fermionic fields).} QFT is sometimes presented as a theory of particles using a Fock space representation.\footnote{See \textcite{schweber_introduction_1961, durr_trajectories_2003, oldofredi_particle_2018, deckert_persistent_2019, durr_understanding_2020, tumulka_foundations_2022}.} In their presentations of spacetime state realism, \textcite{wallace_quantum_2010} and \textcite[sec.\ 8]{wallace_emergent_2012} use a Fock space representation to assign states to regions within non-relativistic quantum physics.  They say little as to why that representation cannot (or should not) be used for relativistic quantum field theory.  Here we consider two ways one might attempt to use a Fock representation to obtain states for regions of space and survey the conceptual and technical problems that each approach runs into, showing that one strategy prevents us from assigning states to regions of space and the other leads to violations of relativistic locality.  We take these problems to give further reasons (beyond those identified elsewhere\footnote{Although we have not seen these problems for a particle approach raised in this way elsewhere, these issues are arguably manifestations of well-known limitations facing particle approaches.  \textcite{malament_defense_1996}, for example, rejects a fundamental ontology of particles in quantum field theory because one cannot simultaneously satisfy four desirable conditions---including a localizability condition (that appears to be violated by the strategy in section \ref{SCOsection}) and a locality condition (that appears to be violated by the strategies in sections \ref{SCOsection} and \ref{NWsection}).  Malament's localizability condition is distinct from our requirement that one be able to assign reduced density matrix states to regions and his locality condition is distinct from the standard of relativistic locality that we introduced in section \ref{EMsection} (and have been applying across theories). Our discussion can be seen as complementing existing literature on localizability and locality by showing explicitly how these problems arise when one attempts to prove that quantum field theory meets the standard for relativistic locality that is applied in electromagnetism.}) to prefer the wave functional approach.  Put another way, these are reasons to prefer a fundamental ontology of fields over particles.  Readers may skip this section if they are already convinced that one should not treat Fock space representations as fundamental within QFT.  To keep things simple and avoid the issues that arise for theories with interactions,\footnote{The Fock space approach runs into problems when one moves beyond the free case discussed here and considers interactions (\cite[sec.\ 4.3]{sebens_fundamentality_2022}). \textcite[pg.\ 847]{fraser_fate_2008} explains that the original Fock space for the non-interacting theory cannot be used for the interacting theory because ``there is no state in the Fock representation for a free field that can be interpreted as containing zero quanta [in the interacting theory]'' (a consequence of Haag's theorem; see also \cite{fraser_particles_2022}).  Put another way, the original Fock representation does not contain the minimum-energy ground state for the Hamiltonian with interactions.  For small interaction terms, the Fock representation may remain useful as an approximation (\cite[pg.\ 280]{wallace_quantum_2022}).  But, it arguably cannot give a fundamental description of the quantum state and thus cannot serve our purpose of proving relativistic locality at the fundamental level within QFT.  Defenders of a fundamental particle approach to QFT have potential ways to save a Fock representation.  \textcite[pg.\ 211]{durr_understanding_2020} appeal to the Dirac Sea.  \textcite[ch.\ 6]{tumulka_foundations_2022} explores using interior-boundary conditions to define Hamiltonians without ultra-violet divergences.} we focus our attention in this section on a free scalar Klein-Gordon field, as in section \ref{WFDsection}.

In a Fock space representation, the universal quantum state is a wave function living in Fock space $\mathcal{F}$, the direct sum of $n$-particle Hilbert spaces $\mathcal{H}$ defined by:
\begin{equation}
    \mathcal{F} = \bigoplus_{N = 0}^{\infty} \mathcal{H}^{(n)}
    \ .
    \label{fockspace}
\end{equation}
These states generalize ordinary quantum mechanical states for a fixed number of particles by allowing for superpositions of different numbers of particles.

We can decompose the quantum state at some time $t$ (adopting the Schr\"odinger picture) in terms of its $n$-particle components, by:
\begin{equation}\label{fockallofspace}
     |\Psi(t) \rangle = |\Psi^{(0)}(t) \rangle + |\Psi^{(1)}(t) \rangle + |\Psi^{(2)}(t) \rangle + ... |\Psi^{(n)}(t) \rangle
     \ .
\end{equation}
Each component can be written in terms of its position space $n$-particle wave function $\psi^{(n)}(\vec{x}_1, ..., \vec{x}_n, t)$ as
\begin{equation}
    |\Psi^{(n)}(t) \rangle = \frac{1}{\sqrt{n!}} \int d^3 \vec{x}_1, ..., d^3 \vec{x}_n \; \psi^{(n)}(\vec{x}_1, ..., \vec{x}_n, t) \hat{a}^\dagger(\vec{x}_1) ... \hat{a}^\dagger(\vec{x}_n) |\Omega\rangle 
    \ ,
    \label{fockspacecomponents}
\end{equation}
where $|\Omega\rangle$ is the vacuum state,  $\hat{a}^\dagger(\vec{x})$ creates a particle at $\vec{x}$, $\frac{1}{\sqrt{2}}\hat{a}^\dagger(\vec{x})\hat{a}^\dagger(\vec{y})$ creates a symmetric superposition of a particle at $\vec{x}$ and another at $\vec{y}$, and so forth. Because we are dealing with a single type of bosonic particle, the wave function $\psi^{(n)}(\vec{x}_1, ..., \vec{x}_n, t)$ in \eqref{fockspacecomponents} must be symmetric under particle permutation.

We now face a choice point as to how the creation operators in \eqref{fockspacecomponents} should be defined.  The first strategy for finding Fock space states for regions follows the standard textbook definitions and the second adopts the Newton-Wigner definitions.

\subsubsection{Option 1: Standard Creation Operators}\label{SCOsection}

The field operator can be written as an integral over operators that create and annihilate particles with different momenta $\vec{p}$,
\footnote{See \textcite[pg.\ 177]{schweber_introduction_1961}; \textcite[pg.\ 21]{peskin_introduction_1995}.}
\begin{equation}\label{standardfieldoperator}
    \hat{\phi}(\vec{x}) = \int \frac{d^3\vec{p}}{(2\pi)^3} \frac{1}{\sqrt{2\mathcal{E}_{\vec{p}}}} \left( e^{-i\vec{p} \cdot \vec{x}} \;\hat{a}(\vec{p}) + e^{i\vec{p} \cdot \vec{x}} \; \hat{a}^\dagger(\vec{p}) \right)
    \ .
\end{equation}
where $\mathcal{E}_{\vec{p}} = \sqrt{|\vec{p}\,|^2 + m^2}$ is the relativistic energy of a particle with mass $m$ and momentum $\vec{p}$.  One can interpret the second part of this expression \eqref{standardfieldoperator} as creating a particle at $\vec{x}$ and the first part as annihilating a particle at $\vec{x}$, yielding the following creation and annihilation operators:
\begin{align}
    \hat{a}^\dagger(\vec{x}) &= \int \frac{d^3\vec{p}}{(2\pi)^3} \frac{1}{\sqrt{2 \mathcal{E}_{\vec{p}}}} \; e^{i\vec{p}\cdot \vec{x}} \; \hat{a}^\dagger(\vec{p})
\nonumber\\
    \hat{a}(\vec{x}) &= \int \frac{d^3\vec{p}}{(2\pi)^3} \frac{1}{\sqrt{2 \mathcal{E}_{\vec{p}}}} \; e^{-i\vec{p}\cdot \vec{x}} \; \hat{a}(\vec{p})
\label{kgposition}
    \ .
\end{align}
From \eqref{standardfieldoperator} and \eqref{kgposition}, it straightforwardly follows that:
\begin{equation}\label{fieldposition}
    \hat{\phi}(\vec{x}) = \hat{a}(\vec{x}) + \hat{a}^\dagger(\vec{x})
    \ .
\end{equation}

A naive first step towards obtaining reduced density matrices for regions would be to decompose the vacuum $|\Omega\rangle$ into a tensor product of vacua for a region $R$ and its complement $\bar{R}$, as $|\Omega\rangle_R \otimes |\Omega\rangle_{\bar{R}}$---thinking that the union of two regions is in the vacuum state just when each region is in its vacuum state. As \textcite[78]{redhead_vacuum_1994} phrases the intuition, ``the global vacuum implies a local vacuum. If there are no particles anywhere in space, then there are no particles present in any local region of space.'' A naive second step would be to assume that each creation operator $\hat{a}^\dagger(\vec{x})$ in \eqref{fockspacecomponents} either acts in $R$ \textit{or} $\bar{R}$ (leaving the other region in its vacuum state), so that we can rewrite each component of the total state like so:
\begin{equation}\label{firsttryfockspaceregionpartition}
\begin{aligned}
        |\Psi^{(n)}(t) \rangle = \frac{1}{\sqrt{n!}} &\sum_{k=0}^n \int_R d^3\vec{x}_1, ..., d^3\vec{x}_k \; \int_{\bar{R}} d^3\vec{x}_{k+1}, ..., d^3\vec{x}_n  \;  \\ 
        &\psi_n(\vec{x}_1, ..., \vec{x}_n, t)   \hat{a}^{\dagger}(\vec{x}_1) \ldots \hat{a}^{\dagger}(\vec{x}_k) |\Omega\rangle_R \otimes \hat{a}^{\dagger}(\vec{x}_{k+1}) \ldots \hat{a}^{\dagger}(\vec{x}_n) |\Omega\rangle_{\bar{R}}
        \ ,
\end{aligned}
\end{equation}
summing over all the possible ways in which $k$ of the $n$ particles might be distributed across $R$ and $n - k$ across $\bar{R}$ (assuming that $\hat{a}^{\dagger}(\vec{x}_1) \ldots \hat{a}^{\dagger}(\vec{x}_k)$ act only on $|\Omega\rangle_R$ and $\hat{a}^{\dagger}(\vec{x}_{k+1}) \ldots \hat{a}^{\dagger}(\vec{x}_n)$ act only on $|\Omega\rangle_{\bar{R}}$).  From \eqref{fockallofspace} and \eqref{firsttryfockspaceregionpartition}, one could form a global density matrix $|\Psi(t) \rangle\langle \Psi(t) |$ and find a reduced density matrix for the region $R$ by tracing over possible possible particle arrangements in $\bar{R}$.

While somewhat elegant, this approach is not viable.  Both the first and second steps above were flagged as naive and are, in fact, incorrect.  The first step assumed that the global vacuum (or zero-particle state) $|\Omega\rangle$ could be written as $|\Omega\rangle_R \otimes |\Omega\rangle_{\bar{R}}$, an unentangled product state of vacua for $R$ and $\bar{R}$.  But, this runs contrary to the well-known fact that in QFTs (even free QFTs) the vacuum is an entangled state.\footnote{The fact that the vacuum is entangled can be seen either by examining vacuum wave functionals (\cite[sec.\ 10.1]{hatfield_quantum_1992}; \cite[34]{huang_quantum_1998}) or via the Reeh-Schlieder theorem (\cite{clifton_superentangled_1998, fleming_reeh-schlieder_2000, halvorson_reeh-schlieder_2001}).}  The second step assumed that $\hat{a}^\dagger(\vec{x})$ is a local operator, acting only at $\vec{x}$ and not elsewhere.  This is not the case.\footnote{See \textcite[sec.\ 3.3]{piazza_volumes_2008, halvorson_reeh-schlieder_2001}.}  As \textcite[sec.\ 4]{fleming_reeh-schlieder_2000} notes, ``unlike the local field $\hat{\phi}$ itself, the positive and negative frequency parts do not commute (with their adjoints) at space-like separation."  Since these are precisely the creation and annihilation operators that we adopted in \eqref{kgposition} and (in the Heisenberg picture) local operators must commute at spacelike separation \eqref{gencausalitycond}, we see that the standard creation and annihilation operators are not local operators.\footnote{A technical way to put it is that these operators cannot be part of the local algebra of observables associated with a spatial region \parencite[sec.\ 3.1, point 3]{halvorson_reeh-schlieder_2001}.} \textcite{schweber_introduction_1961} explicitly writes down the commutator for $\hat{a}(x)$ and $\hat{a}^\dagger(y)$ in terms of a function that falls off exponentially with $x-y$ but does not vanish for space-like separated $x, y$, showing that they in general do not commute.\footnote{A similar derivation can be found in \textcite[pg.\ 44]{henley_elementary_1962}.  See also \textcite[sec.\ 6.5]{duncan_conceptual_2012}.}

Noting the failure of the first step and the entanglement of the vacuum, one might back up and attempt to write the global vacuum state as an entangled state of $R$ and $\bar{R}$.  Let us suppose that the global Hilbert space \eqref{fockspace} (Fock space) can be factorized into separate Hilbert spaces for $R$ and $\bar{R}$, $\mathcal{F} = \mathcal{H}_{R} \otimes \mathcal{H}_{\bar{R}}$.\footnote{In full QFT, analyzed at a high standard of mathematical rigor, this decomposition may not exist (\cite[pg.\ 942--943]{swanson_how_2020}; \cite[footnote 3]{falcone_localization_2024}).  However, in section \ref{WFFSsection} we seemed to be able to make such a decomposition.  Also, \textcite{wallace_quantum_2010, wallace_emergent_2012} appeal to such a decomposition when they adopt an algebraic perspective for assigning states to spacetime regions.  For our purposes here, we will help ourselves to this decomposition.}  Then, Schmidt decomposition lets us write the global vacuum, $|\Omega\rangle$, as a sum over tensor product states of $\mathcal{H}_R$ and $\mathcal{H}_{\bar{R}}$:
\begin{equation}\label{zeroright}
    |\Omega\rangle = \sum_{i, j} \lambda_{ij} |\Omega_i\rangle_R \otimes |\Omega_j\rangle_{\bar{R}}
    \ ,
\end{equation}
where $|\Omega_i\rangle_R$ are orthogonal states of $\mathcal{H}_R$, $|\Omega_j\rangle_{\bar{R}}$ orthogonal states of $\mathcal{H}_{\bar{R}}$, and $\lambda_{ij}$ are real-valued Schmidt coefficients for each product state.

The naive approach to factorizing the vacuum as an unentangled product state of vacua for $R$ and $\bar{R}$ used earlier, $|\Omega\rangle=|\Omega\rangle_R \otimes |\Omega\rangle_{\bar{R}}$, came with a clear physical meaning: the world as a whole has no particles and each region is also in a vacuum state. The Schmidt decomposition in \eqref{zeroright} jettisons this physical picture and it is unclear how the superposed states $|\Omega_i\rangle_R$ and $|\Omega_j\rangle_{\bar{R}}$ are to be interpreted. Two ways forward both lead to interpretive puzzles. 

First, a natural interpretation of the states of these regional Hilbert spaces would be in terms of particle numbers associated with each region. This interpretation forces us to say (absurdly) that the global vacuum state (with no particles) is a superposition of states where there are various different numbers of particles in $R$ and $\bar{R}$.  As \textcite[pg.\ 11]{falcone_localization_2024} put it, ``the vacuum is not locally devoid of quanta, but only globally.''  Second, we might view $|\Omega_i\rangle$ and $|\Omega_j\rangle$ as distinct vacuum states for $R$ and $\bar{R}$ that describe a variety of different zero-particle states for those regions.  What then is the distinction between these states and how can it be captured in the particle language that accompanies the Fock space approach? For our purposes here, let us press on, having flagged these interpretive issues.

A general state could be constructed from \eqref{zeroright} by acting with creation operators on this vacuum, as in \eqref{fockallofspace} and \eqref{fockspacecomponents}.  However, because of the failure of the second step taken earlier, one cannot assume that each creation operator $\hat{a}^\dagger(\vec{x})$ acts only on either the $R$ or $\bar{R}$ states in the Schmidt decomposition \eqref{zeroright} (depending on where $\vec{x}$ is in $R$ or $\bar{R}$), as was assumed to get \eqref{firsttryfockspaceregionpartition}.  This prevents us from defining the reduced density matrix for $R$ by tracing over possible particle arrangements in $\bar{R}$, as was suggested under \eqref{firsttryfockspaceregionpartition}.

We have thus seen that the strategy for assigning reduced density matrix states to regions outlined earlier---in the paragraph that includes \eqref{firsttryfockspaceregionpartition}---fails at multiple points and cannot be easily patched up.  One could try a radically different approach.\footnote{We thank David Baker for this suggestion.}  At least in the context of non-interacting QFTs, there are recipes for writing particle wave functions (Fock space states) as field wave functionals.\footnote{See \textcite[sec.\ 10.1]{hatfield_quantum_1992}; \textcite[sec.\ 11.5]{bohm_undivided_1993}; \textcite{baker_against_2009}; \textcite[sec.\ 4.3]{sebens_fundamentality_2022}.}  As we have seen that field wave functionals do yield reduced density matrix states for regions, one might use the following two-step procedure to find states for a region $R$: first (i) rewrite the global Fock space state \eqref{fockallofspace} as a global wave functional, and then (ii) use the method from section \ref{WFFSsection} to find a reduced density matrix state for $R$.  The problem with this strategy is that it is parasitic on the field wave functional approach and the reduced density matrices that you end up with are not immediately interpretable in terms of particles.  These are not viable \emph{particle} states for regions.  If one is going to use the field reduced density matrices from section \ref{WFFSsection} as states for regions, then it seems like one should use wave functionals to represent global states and take a thoroughgoing field approach to quantum field theory (as in sections \ref{WFFSsection} and \ref{WFDsection}).

To attempt a proof of relativistic locality (or its violation), we need to start with states for regions.  Unfortunately, we have not been able to write down a general expression for such states within a particle approach to QFT (using the standard creation and annihilation operators).  Thus, we reach the inconclusive conclusion that we cannot determine whether this Fock space version of QFT satisfies the standard of relativistic locality from section \ref{EMsection} because the non-locality of the particle creation operators has prevented us from finding a way to assign states to regions.

\subsubsection{Option 2: Newton-Wigner Creation Operators}\label{NWsection}

Let us now back up and consider an alternative proposal as to how we ought to define the creation operators that appear in \eqref{fockspacecomponents}, a proposal that has been advocated by \textcite{pagonis_strange_1999, fleming_reeh-schlieder_2000}. Removing the factors of $\frac{1}{\sqrt{2 \mathcal{E}_{\vec{p}}}}$ that appear in \eqref{kgposition},\footnote{This seemingly inconsequential move breaks the Lorentz covariance of the measure, which leads to the drastic physical differences for the Newton-Wigner proposal we see below.} we can put forward new creation operators for particles at specific positions (written in terms of the same old creation operators for particles with specific momenta):
\begin{align}
    \hat{a}_{NW}^\dagger(\vec{x}) &= \int \frac{d^3\vec{p}}{(2\pi)^3} \; e^{i\vec{p}\cdot \vec{x}} \; \hat{a}^\dagger(\vec{p})
\nonumber\\
    \hat{a}_{NW}(\vec{x}) &= \int \frac{d^3\vec{p}}{(2\pi)^3} \; e^{-i\vec{p}\cdot \vec{x}} \; \hat{a}(\vec{p})
\label{Nwcreationannihilation}
    \ .
\end{align}
These are the Newton-Wigner creation and annihilation operators.\footnote{See \textcite[pg.\ 9--10]{fleming_reeh-schlieder_2000}; \textcite[sec.\ 2]{piazza_volumes_2008}; \textcite[sec.\ 2.2]{falcone_localization_2024}.}

One can introduce a different way of carving the global Hilbert space (Fock space) into Hilbert spaces for regions such that the Newton-Wigner creation and annihilation operators for points in a region act only on that region's Hilbert space.  Relative to this revised carving, it is possible to regard the vacuum as an unentangled state\footnote{For technical details, \textcite[sec.\ 4--5]{fleming_reeh-schlieder_2000}; \textcite[sec.\ 4]{halvorson_reeh-schlieder_2001}; \textcite{piazza_volumes_2008}; \textcite[sec.\ 5.1.2]{falcone_localization_2024}.} such that
\begin{equation}
    |\Omega\rangle = |\Omega_{NW}\rangle_R \otimes |\Omega_{NW}\rangle_{\bar{R}}
    \ .
    \label{vacuumproductstate}
\end{equation}
The Newton-Wigner approach thus seems to capture the intuitive physical meaning we sought earlier, whereby ``the vacuum \dots is devoid of quanta locally as well as globally" \parencite[pg.\ 11]{fleming_reeh-schlieder_2000}.

We can proceed to build the $n$-particle piece \eqref{fockspacecomponents} of a general state \eqref{fockallofspace} by acting on the vacuum with Newton-Wigner creation operators, operators that will act either on $|\Omega_{NW}\rangle_R$ (if $\vec{x}$ is in $R$) or $|\Omega_{NW}\rangle_{\bar{R}}$ (if $\vec{x}$ is in $\bar{R}$),
\begin{equation}\label{secondtryfockspaceregionpartition}
\begin{aligned}
        |\Psi^{(n)}(t) \rangle = \frac{1}{\sqrt{n!}} &\sum_{k=0}^n \int_R d^3\vec{x}_1, ..., d^3\vec{x}_k \; \int_{\bar{R}} d^3\vec{x}_{k+1}, ..., d^3\vec{x}_n  \;  \\ 
        &\psi_n(\vec{x}_1, ..., \vec{x}_n, t)   \hat{a}_{NW}^{\dagger}(\vec{x}_1) \ldots \hat{a}_{NW}^{\dagger}(\vec{x}_k) |\Omega_{NW}\rangle_R \;  \\ 
        &\otimes \hat{a}_{NW}^{\dagger}(\vec{x}_{k+1}) \ldots \hat{a}_{NW}^{\dagger}(\vec{x}_n) |\Omega_{NW}\rangle_{\bar{R}}
        \ ,
\end{aligned}
\end{equation}
This takes the form that we hoped for in \eqref{firsttryfockspaceregionpartition}, which was blocked by the entanglement of the vacuum (now solved) and the non-locality of the standard creation operators. 

The Newton-Wigner creation operators are spatially local, in the sense that creating a particle at a moment has no effect elsewhere, allowing us to write \eqref{secondtryfockspaceregionpartition}.  The Schr\"{o}dinger-picture creation and annihilation operators commute at spatial separation (\cite[sec.\ 4]{fleming_reeh-schlieder_2000}; \cite[sec.\ 5.1.2]{falcone_localization_2024}).  The Heisenberg-picture creation and annihilation operators obey equal-time commutation relations, but do not commute at space-like separation (\cite[sec.\ 5]{halvorson_reeh-schlieder_2001}; \cite[pg.\ 5]{piazza_volumes_2008}).  That failure to commute at space-like separation means that the Heisenberg-picture operators are not local operators. This leads to violations of relativistic locality, as we will show.

Within the Newton-Wigner approach, one can construct reduced density matrix states for regions.  To see how this is done, let us begin by introducing a more compact notation for \eqref{secondtryfockspaceregionpartition},
\begin{equation}\label{secondtryfockspaceregionpartition2}
\begin{aligned}
        |\Psi^{(n)}(t) \rangle = \frac{1}{\sqrt{n!}} &\sum_{k=0}^n \int_R d^3\vec{x}_1, ..., d^3\vec{x}_k \; \int_{\bar{R}} d^3\vec{x}_{k+1}, ..., d^3\vec{x}_n  \;  \\ 
        &\psi^{(n)}(\vec{x}_1, ..., \vec{x}_n, t) |\vec{x}_1, ..., \vec{x}_k \rangle_R \otimes |\vec{x}_{k+1}, ..., \vec{x}_n \rangle_{\bar{R}}
        \ ,
\end{aligned}
\end{equation}
where the creation operators have been absorbed into the states of $R$ and $\bar{R}$.  From \eqref{fockallofspace}, it is clear that the universal density matrix at time $t$, $\hat{\rho}(t)$, is simply:
\begin{equation}
    \hat{\rho}(t) = |\Psi(t) \rangle \langle \Psi(t) | = \sum_{n, m} |\Psi^{(n)}(t) \rangle \langle \Psi^{(m)}(t) | = \sum_{n, m} \rho_{nm}(t)
\end{equation}
To find the $n,m$-components of the universal density matrix, $\rho_{nm}(t)$, we can use \eqref{secondtryfockspaceregionpartition2}:
\begin{equation}
    \begin{aligned}
    \hat{\rho}_{nm}(t) & = |\Psi^{(n)}(t) \rangle \langle \Psi^{(m)}(t) | 
    \\ & = \frac{1}{\sqrt{n!}\sqrt{m!}} \sum_{k=0}^n  \sum_{j=0}^{m} \int_R d^3\vec{x}_1 ... d^3\vec{x}_k \int_{\bar{R}} d^3\vec{x}_{k+1} ... d^3\vec{x}_n \int_R d^3\vec{x}'_1 ... d^3\vec{x}'_j \int_{\bar{R}} d^3\vec{x}'_{j+1} ... d^3\vec{x}'_{m} \\
    & \psi^{(n)}(\vec{x}_1, ..., \vec{x}_n, t) \psi^{*(m)}(\vec{x}'_1, ..., \vec{x}'_{m}, t) \\ & |\vec{x}_1, ..., \vec{x}_k \rangle_R \otimes |\vec{x}_{k+1}, ..., \vec{x}_n \rangle_{\bar{R}} \langle \vec{x}'_1, ..., \vec{x}'_j |_R \otimes \langle \vec{x}'_{j+1}, ..., \vec{x}'_n |_{\bar{R}}
    \ .
    \end{aligned}
\end{equation}
To find the reduced density matrix for $R$, $\hat{\rho}_R$, we perform the partial trace over $\bar{R}$,
\begin{equation}
    \hat{\rho}_R(t) = \mbox{tr}_{\bar{R}}\left(\hat{\rho}(t)\right)= \sum_{n} \sum_{m} \mbox{tr}_{\bar{R}}\left(\hat{\rho}_{nm}(t)\right) \,.
    \label{nmdecomp}
\end{equation}
We can trace over $\bar{R}$ by summing over the set of orthonormal states $\{|0\rangle_{\bar{R}}, |\vec{x}_1 \rangle_{\bar{R}}, |\vec{x}_1, \vec{x}_2\rangle_{\bar{R}}, ..., |\vec{x}_1, ..., \vec{x}_n\rangle_{\bar{R}}\}$. Thus, for each component of the density matrix $\rho_{nm}(t)$, we can find its contribution to the reduced density matrix for region $R$ as follows:
\begin{equation}\label{Focklocalstate}
\begin{aligned}
       \mbox{tr}_{\bar{R}}\left(\hat{\rho}_{nm}(t)\right) & = \frac{1}{\sqrt{n!}\sqrt{m!}} \sum_{k=0}^n  \sum_{j=0}^{m}\int_R d^3\vec{x}_1 ... d^3\vec{x}_k \int_{\bar{R}} d^3\vec{x}_{k+1} ... d^3\vec{x}_n \int_R d^3\vec{x}'_1 ... d^3\vec{x}'_j \int_{\bar{R}} d^3\vec{x}'_{j+1} ... d^3\vec{x}'_{m} \\
        & \psi^{(n)}(\vec{x}_1, ..., \vec{x}_n, t) \psi^{*(m)}(\vec{x}'_1, ..., \vec{x}'_{m}, t) \\ &  |\vec{x}_1, ..., \vec{x}_k \rangle_R \langle \vec{x}'_1, ..., \vec{x}'_j |_R  \langle \vec{x}_{k+1}, ..., \vec{x}_n |\vec{x}'_{j+1}, ..., \vec{x}'_{m} \rangle_{\bar{R}}  \\
        & = \frac{1}{\sqrt{n!}\sqrt{m!}}\sum_{l=0}^{n\ \mbox{\scriptsize{or}}\ m} l!\int_R d^3\vec{x}_1 ... d^3\vec{x}_{n-l} \int_{\bar{R}} d^3\vec{x}_{n-l+1} ... d^3\vec{x}_n \int_R d^3\vec{x}'_1 ... d^3\vec{x}'_{m-l} \\
        & \psi^{(n)}(\vec{x}_1, ..., \vec{x}_n, t) \psi^{*(m)}(\vec{x}'_1, ..., \vec{x}'_{m-l}, \vec{x}_{n-l+1}, ..., \vec{x}_{n}, t) \\ &  |\vec{x}_1, ..., \vec{x}_{n-l} \rangle_R \langle \vec{x}'_1, ..., \vec{x}'_{m-l} |_R  
\end{aligned}
\end{equation}
where only the terms where there are the same number of particles outside $R$ ($m - j = n - k$) survive from the first line to the second (due to orthogonality considerations).  This count is denoted by $l$ and summed from zero to $n$ or $m$, whichever is smaller, in the second line.  The factor of $l!$ is a combinatorial factor that arises because of the different ways that the particles outside of $R$ might be paired up.  In \eqref{Focklocalstate}, we have assumed that each $\psi^{(n)}$ is symmetric under particle permutations (as must be the case for bosonic wave functions). Together, \eqref{nmdecomp} and \eqref{Focklocalstate} give the general form of Newton-Wigner Fock states for regions of space at a time (something that we have not seen presented elsewhere).

To better understand \eqref{Focklocalstate} and to set up our discussion of locality, let us derive the reduced density matrix state for a region $R$ when the global state is a general single-particle state (where the only contribution to \eqref{fockallofspace} is $|\Psi^{(1)}(t) \rangle$):
\begin{equation}
    |\Psi(t) \rangle =  \int d^3 \vec{x}\; \psi(\vec{x}, t) \; \hat{a}_{NW}^\dagger(\vec{x})|\Omega_{NW} \rangle 
    \ .
    \label{fockspace1particleNW}
\end{equation}
Applying \eqref{Focklocalstate} with $n=m=1$, the reduced density matrix for $R$ at $t$ is
\begin{align}\label{justinRfock}
    \hat{\rho}^{R}(t) &=
    \int_{R} d^3\vec{x} \; d^3\vec{x}'\;\psi(\vec{x}, t)\psi^*(\vec{x}', t) \;|\vec{x} \rangle_{R}\langle \vec{x}' |_{R}
    \nonumber
    \\
     &\quad+\int_{\bar{R}} d^3\vec{x}\;\psi(\vec{x}, t)\psi^*(\vec{x}, t)\; |\Omega_{NW} \rangle_{R}\langle \Omega_{NW} |_{R}
     \ ,
\end{align}
an expression that sums a contribution for the particle being within $R$ and a contribution for it being outside of the region. 

With states for regions in hand, we can now discuss relativistic locality. \textcite[pg. 504]{fleming_reeh-schlieder_2000} observes that ``\textit{NW} localized states have a superluminal contribution to their evolution,''\footnote{Other authors have also concluded that the Newton-Wigner approach leads to non-locality (\cite[pg.\ 55]{fulling_aspects_1989}; \cite[pg.\ 4]{piazza_volumes_2008}; \cite{pavsic_localized_2018}; \cite[pg.\ 2]{falcone_localization_2024}).} citing the fact that (using Heisenberg picture operators)
\begin{equation}
    \langle\Omega| \hat{a}_{NW}(\vec{x},t)\hat{a}_{NW}^{\dagger}(\vec{x}',t')|\Omega\rangle \neq 0
    \label{superluminaljump}
\end{equation}
even when $(\vec{x},t)$ and $(\vec{x}',t')$ are space-like separated, something that would not be true if these operators commuted at space-like separation. That is, the Newton-Wigner creation operators violate \eqref{gencausalitycond} and thus are not local operators.

We can readily see how \eqref{superluminaljump} leads to a violation of the standard for relativistic locality from section \ref{EMsection}. Revisiting figures \ref{lightcones} and \ref{frustumfig}, we will show that two global states can initially agree within $R$ and later disagree within $R^-$, because initial differences in $\bar{R}$ can affect what later occurs within $R^-$, \eqref{fockspace1particleNW} can be expressed using Heisenberg picture creation operators as
\begin{equation}
    |\Psi(t) \rangle =  \int d^3 \vec{x}\; \psi(\vec{x}) \hat{a}_{NW}^\dagger(\vec{x},t)|\Omega\rangle 
    \ .
    \label{fockspace1particleNWHP}
\end{equation}
Suppose that $\psi(\vec{x})$ is zero within $R$.  Then, the initial state can be written as
\begin{equation}
    |\Psi(0) \rangle =  \int_{\bar{R}} d^3 \vec{x}\; \psi(\vec{x}) \; |\Omega_{NW}\rangle_R \otimes \hat{a}_{NW}^\dagger(\vec{x},0) |\Omega_{NW}\rangle_{\bar{R}}
    \ ,
    \label{}
\end{equation}
where $\hat{a}_{NW}^\dagger(\vec{x},0)$ acts only on the vacuum for $\bar{R}$ and not the vacuum for $R$.  At a later time $t$,
\begin{equation}
    |\Psi(t) \rangle \neq  \int_{\bar{R}} d^3 \vec{x}\; \psi(\vec{x}) \; |\Omega_{NW}\rangle_{R^-} \otimes \hat{a}_{NW}^\dagger(\vec{x},t) |\Omega_{NW}\rangle_{\bar{R}^-}
    \ .
    \label{}
\end{equation}
That expression is not correct because, for $\vec{x}\in \bar{R}$, $\hat{a}_{NW}^\dagger(\vec{x},t)|\Omega\rangle$ is not the product of the vacuum within $R^-$ with some single-particle state confined to $\bar{R}^-$. (If it were, then inner products of the form $\langle\Omega| \hat{a}_{NW}(\vec{y}, t)\hat{a}_{NW}^{\dagger}(\vec{x},0)|\Omega\rangle$ would always be zero for $\vec{x}\in \bar{R}$ and $\vec{y}\in R^-$, contra \eqref{superluminaljump}.)  The fact that $\hat{a}_{NW}^\dagger(\vec{x},t)$ alters the state within $R^-$ even when $\vec{x}$ is in $\bar{R}$ --- the fact that $\hat{a}_{NW}^\dagger(\vec{x},t)$ is not a local operator --- is what leads to the violation of relativistic locality.  Returning to our standard for relativistic locality from section \ref{EMsection}: two single particle states \eqref{fockspace1particleNWHP} that initially agree on $\psi(\vec{x})$ within $R$, and thus agree on the reduced density matrix within $R$ \eqref{justinRfock}, can later disagree on the reduced density matrix within $R^-$ because differences in the factors $\psi(\vec{x})$ preceding $\hat{a}_{NW}^\dagger(\vec{x},t)|\Omega\rangle$ for $\vec{x}$ in $\bar{R}$ will lead to differences within $R^-$.

We have thus seen that a particle approach to QFT leads to non-locality if we use the Newton-Wigner creation operators.  In section \ref{SCOsection}, we saw that the standard creation operators do not allow one to assess the locality of the dynamics because they do not yield a way of assigning states to regions.  All of this illustrates a significant advantage of the field approach over a particle approach.\footnote{Some other advantages are enumerated in \textcite{sebens_fundamentality_2022}.} Only in terms of wave functionals, rather than particle states, do we see relativistic locality.  Making use of the fact that the Heisenberg-picture field operators are local operators that commute at space-like separation (unlike the standard or Newton-Wigner particle creation operators), one can show that the wave functional approach satisfies relativistic locality (section \ref{WFDsection}).  It turns out that whether QFT is relativistically local or not depends on whether one uses particles or fields to articulate the fundamental laws and ontology of the theory.

\section{Conclusion}

In this paper we have argued that the unitary evolution of the quantum state in QFT without collapse (Everettian QFT) is local.  We began by putting forward a single standard for relativistic locality that can be applied across different theories and then showed explicitly that electromagnetism, the Klein-Gordon equation, and the Dirac equation meet that standard.  For QFT, one can debate whether it is better to take a field (wave functional) or a particle (Fock space) approach.  Here we have focused our attention on one advantage of the field approach, showing (in a rough way that could be made more rigorous) that a field approach to QFT achieves relativistic locality whereas a particle approach to QFT is either non-local or fails to assign states to regions.

Throughout our discussion of different theories, we have focused on the evolution of the fundamental ontology and not on any interventions or measurements that might be conducted by observers.  If the underlying dynamics are local, such interventions cannot violate that locality.

We thus take the locality of the many-worlds interpretation to be established by studying the fundamental ontology and dynamics, as was done in section \ref{QFTsection}. The many-worlds interpretation posits a universal quantum state (a wave function, wave functional, or density matrix\footnote{See \textcite{Chuachen2025}.}) that evolves under unitary dynamics (without collapse).  The theory does not have additional postulates dividing this quantum state into worlds or saying precisely when one world branches into many.  Worlds are non-fundamental entities that emerge (via decoherence) as convenient ways of carving the quantum state into approximately non-interacting pieces.  This is why \textcite{wallace_emergent_2012} titled his book \emph{The Emergent Multiverse}.  Because worlds are merely an emergent higher-level description of the fundamental ontology, we do not see debates about how the wave function should be carved into worlds as challenging the locality of the many-worlds interpretation.

\textcite[ch.\ 8]{wallace_emergent_2012} advocates a local branching view, according to which quantum measurements cause a local division of the one world into many, with branching propagating no faster than the speed of light.  By contrast, \textcite{sebens_self-locating_2018} argue that one can take a global branching view where a quantum measurement anywhere causes the entire world (and everything in it) to branch.  This global picture of branching plays a role in the Sebens-Carroll derivation of the Born Rule and has been challenged by \textcite{kent_does_2015, mcqueen_defence_2019} (and defended by Ney, \citelink{ney_argument_nodate}{forthcoming}, and Ney, \citelink{ney_branching_nodate}{this volume}).\footnote{Ney (\citelink{ney_argument_nodate}{forthcoming}) gives a detailed analysis of the debate between global and local branching.  She also discusses a third option, semi-local branching, that she attributes to \textcite{mcqueen_defence_2019}. \textcite{gao_understanding_2024} gives a fourth option that he calls ``nonlocal branching.''}

Electrostatics is a non-local theory that emerges as an approximation to our paradigm case of a local theory: electromagnetism.  Newtonian gravity is a non-local theory that emerges as an approximation to another paradigm case of a local theory: general relativity.  These two examples demonstrate that it is perfectly fine to have non-fundamental non-locality emerge from underlying local physics.  For this reason, we see the debate about the locality of branching as orthogonal to our focus in this paper on fundamental locality.  Because the fundamental dynamics of Everettian QFT are local, the many-worlds interpretation is local.

\section*{Acknowledgments}
We would like to thank David Baker, Jacob Barandes, Alexander Blum, Francisco Calder\'{o}n, Mario Hubert, Nick Huggett, Dustin Lazarovici, Alyssa Ney, Noel Swanson, Roderich Tumulka, and David Wallace for helpful comments and discussions.

\nocite{adlam_should_nodate}
\nocite{vaidman_many-worlds_nodate}
\nocite{ney_argument_2025}
\nocite{ney_branching_nodate}

\printbibliography

\end{document}